\documentclass[aps,prc,preprint,groupedaddress,nofootinbib,showpacs,preprintnumbers,amsmath,amssymb,superscriptaddress]{revtex4} 
 
\usepackage{graphicx}
\usepackage{dcolumn}
\usepackage{bm}
\usepackage{amsmath}

\begin{document} 
 
\def\R{\right} \def\L{\left} \def\Sp{\quad} \def\Sp2{\qquad} 
 
 

\title{Quasifree eta photoproduction from nuclei and medium
modifications of resonances} 


\author{B.I.S. van der Ventel}
\email{bventel@sun.ac.za}
\affiliation{Department of Physics, University of Stellenbosch,
Stellenbosch 7600, South Africa}

\author{L. J. Abu-Raddad} 
\affiliation{Research Center for Nuclear Physics, Osaka University, 10-1 Mihogaoka, Ibaraki, Osaka 567-0047, Japan}  
\affiliation{Department of Infectious Disease Epidemiology,
Imperial College Faculty of Medicine, St Mary's Campus, Norfolk Place,
London W2 1PG, United Kingdom} 

\author{G.C. Hillhouse}
\affiliation{Department of Physics, University of Stellenbosch,
Stellenbosch 7600, South Africa}
\affiliation{Research Center for Nuclear Physics, Osaka University, 10-1 Mihogaoka, Ibaraki, Osaka 567-0047, Japan}  
\date{\today} 
 
\begin{abstract} 
This article establishes the case that the process of quasifree $\eta$
photoproduction from nuclei is a prized tool to study medium
modifications and changes to the elementary process $\gamma N
\rightarrow \eta N$ in the nuclear medium. We investigate the
sensitivity of the differential cross section, recoil nucleon
polarization and the photon asymmetry to changes in the elementary
amplitude, medium modifications of the resonance $(S_{11},D_{13} )$
masses, as well as nuclear target effects.  All calculations are
performed within a relativistic plane wave impulse approximation
formalism resulting in analytical expressions for all observables. Our
results indicate that polarization observables are largely insensitive
to nuclear target effects. Depending on the type of coupling, the spin
observables do display a sensitivity to the magnitude of the $\eta NN$
coupling constant.  The polarization observables are identified to be
the prime candidates to investigate the background processes and their
medium modifications in the elementary process such as the $D_{13}$
resonance. Moreover, as a consequence of the large dominance in the
differential cross section of the $S_{11}$ resonance, the quasifree
differential cross section provides an exceptional instrument to study
medium modifications to the $S_{11}$ resonance in such a manner that
helps to distinguish between various models that attempt to understand
the $S_{11}$ resonance and its distinctive position as the lowest
lying negative parity state in the baryon spectrum.

\end{abstract} 
 

\maketitle 
\section{Introduction} 
\label{intro}

The $\eta$ electro- and photoproduction processes continue to enjoy 
significant investigations from a variety of approaches. This 
interest stems from the fact that $\eta$ processes form a gate to 
understand several fundamental puzzling issues in nuclear and 
particle physics today such as measuring the $\bar{s}s$ quark content in the
nucleon~\cite{DF90}. While the $\eta$ photoproduction process is 
only one of many meson photoproduction processes from nuclei, the characteristic 
reactive content of this process near threshold provides it with a 
distinguished role among other meson photoproduction processes. Part of 
the reason is that the interaction is dominated cleanly by only one 
resonance near threshold. This is the $S_{11}$(1535) resonance with its 
peculiar 
status as the lowest lying negative parity resonance in the baryon 
spectrum. Favorably, the theoretical interest has been correlated with 
experimental advances, particularly with the construction and running of modern electron-scattering  
facilities, such as the Thomas Jefferson National Accelerator  
Facility (JLab) and Mainz~\cite{krus95a,krus95b}.  
 
Understanding the structure of the $S_{11}$ resonance is of prime 
interest in baryon physics as it is believed that this resonance plays a crucial role in 
the dynamics of chiral symmetry and its spontaneous breaking in the 
baryon spectrum. The $S_{11}$ resonance has been studied using various 
approaches including  
effective Lagrangian theory~\cite{BMN91,Li95,bmz95,KNS95,SFN95} and QCD 
sum rules~\cite{JKO96,KL97,LK97,JOH98}. Yet, a more unifying 
approach is to study this resonance using chiral symmetry and its 
spontaneous breaking and restoration. Indeed, DeTar and Kunihiro have suggested the ``mirror 
assignment'' of chiral symmetry where the $S_{11}$ resonance, as the lowest lying 
negative parity resonance, is the chiral 
partner of the nucleon and transforms in an opposite direction compared 
to the nucleon~\cite{DK89}. However, Jido, Nemoto, Oka, and Hosaka have suggested 
a different realization of chiral symmetry through what they label 
as the ``na\"{i}ve assignment''~\cite{JNOH00,JOH01}. Improvements to these models have 
been suggested through the inclusion of nonlinear terms that 
preserve chiral symmetry~\cite{KJO98}. Understanding the medium 
modifications of the $S_{11}$ resonance will help us discriminate between the 
different models of chiral symmetry assignment, and in turn, 
this leads us to understanding the structure of this resonance and 
the prediction of its properties under chiral symmetry restoration.

It had been thought earlier that the coherent $\eta$ photoproduction 
process from nuclei may in fact help us to study the medium 
modifications of the $S_{11}$ resonance~\cite{oka}. Nonetheless, it has been shown 
that the $S_{11}$ contribution, although dominant for the elementary 
process $\gamma N \rightarrow \eta N$, is strongly suppressed in the 
coherent process due to the filtering of only the isoscalar 
contribution and due to the spin-flip nature of the $S_{11}$ exchange 
diagram~\cite{PSB97,PLM98,APSR99}. Hence, the $\eta$ process $A(\gamma , \eta 
N)B$ in the 
quasifree regime, with its strong dependence on the resonance 
contribution, provides an opportunity to understand the medium 
modifications of the $S_{11}$ resonance.

Building on a series of $\eta$ 
photoproduction studies~\cite{bt90,HR90,Carrasco93,CC94,TBK94,BB94,HETM95}, 
Lee, Wright, Bennhold, and Tiator studied the $\eta$ quasifree 
process using the nonrelativistic distorted wave impulse approximation 
(DWIA) formalism~\cite{lwbt96}. In this work, we also assume the impulse 
approximation but provide a 
fully relativistic study in both the reactive 
content and the nuclear structure. Furthermore,  
we use a different and robust dynamical 
content for the elementary process~\cite{benm92,mzb95,bmz95} and study
this process in a different kinematic setting compared 
to the one used by Lee {\it et al}. Our article constitutes the third application of our established quasifree formalism~\cite{thesis} 
after studying the kaon~\cite{AP00} and the electron~\cite{AP01} quasifree 
reactions.  The goal here 
is to shed light on the elementary process $\gamma N \rightarrow \eta 
N$ by furnishing a different physical setting from the on-shell point 
for studying the elementary amplitude. We also examine the possibility 
of using this process to extract medium modification effects to the 
propagation of the $S_{11}$ and $D_{13}$ resonances. We provide
special attention in our work to the 
polarization observables, the recoil polarization of the ejected 
nucleon and the photon asymmetry, as they are very sensitive to the 
fine details in the reactive content and are effective discriminators of 
subtle physical effects compared to the unpolarized differential cross 
section. Moreover, the quasifree polarization observables, while very 
sensitive to the fundamental processes, are insensitive to distortion 
effects~\cite{lwb93,lwbt96,blmw98}. Finally, by comparing against the 
polarization observables of the free process, medium effects can be discerned.

The effects of relativity in meson quasifree processes are still 
not well understood. While it appears that the polarization observables 
are not sensitive to the enhanced lower component of the Dirac spinors 
in the relativistic nuclear structure 
models~\cite{AP00}, other results indicate important differences compared to
nonrelativistic Schr\"{o}dinger-based formalisms. 
Relativistic plane-wave impulse 
approximation (RPWIA) calculations, such as the one we present in this 
study, have identified subtle dynamics not present at the 
nonrelativistic level. Two notable examples to quote here are that relativistic 
effects contaminate any attempt to infer color transparency from 
a measurement of the asymmetry in the $(e,e'p)$ reaction~\cite{gp94} 
and the breakdown of the nonrelativistic factorization picture in the 
$(e,e'p)$ process as a consequence of the presence of negative-energy 
components in the bound-nucleon wave function \cite{cdmu98}.

Although processes of $\eta$ photoproduction from nucleons and nuclei
have recently enjoyed significant experimental attention, the current experimental work
for quasifree scattering has thus far been limited to the measurement of the
total and differential cross sections. This includes scattering off the 
deuteron~\cite{krus95b} and $^{4}$He~\cite{Hejny99,Weiss_EJoPA13_2001} as well as heavier
nuclei~\cite{RobigLandau_PLB373_1996}. Nonetheless what is
theoretically desired are measurements of the polarization observables
as these observables continue to show the promise of discerning the
subtle dynamics in meson photoproduction processes. With the advent of
JLab and Mainz such measurements are finally realizable.

We have organized our paper as follows. In Sec.~\ref{sec:formal} we 
discuss our relativistic plane-wave formalism, placing 
special emphasis on the use of the ``bound-nucleon propagator'' to 
enormously simplify the formalism. 
In Sec.~\ref{sec:results}, this formalism is applied to calculate the
unpolarized differential cross section, the recoil nucleon polarization and the
photon asymmetry. Finally, conclusions are drawn in Sec.~\ref{sec:concl}.

\section{Formalism} 
\label{sec:formal} 
 
\begin{figure} 
\includegraphics[totalheight=2.5in,angle=0]{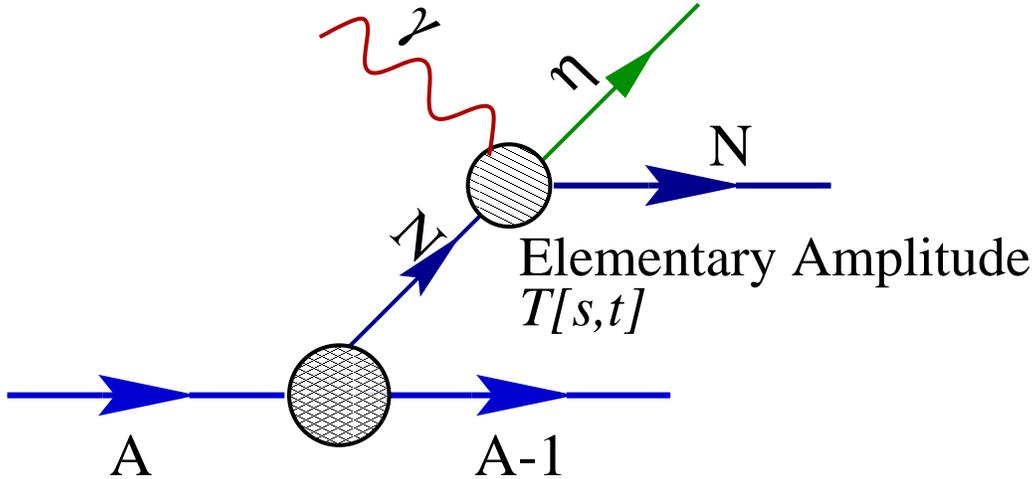} 
 \caption{\label{fig1} A schematic diagram of the $\eta$ meson 
quasifree photoproduction within the framework of a plane-wave  
	  impulse-approximation approach.} 
\end{figure}

\subsection{Kinematic constraints} 
\label{subsec:kin} 
 
In the meson quasifree photoproduction $A(\gamma,\eta N)B$, 
the kinematics are constrained by two conditions. The first is an overall  
energy-momentum conservation: 
\begin{equation} 
k + p_A = k^{\prime} + p^{\,\prime} + p_B \;. 
\end{equation} 
Here $k$ is the four-momentum of the incident photon, while 
$k^{\prime}$ and $p^{\prime}$ are the momenta of the produced $\eta$  
and nucleon, respectively. $p_A$($p_B$) represents  
the momentum of the target (recoil) nucleus. Since we assume 
the impulse approximation, as shown in 
Fig.~\ref{fig1}, another 
kinematical constraint emerges from the energy-momentum conservation at 
the $\gamma N \!\rightarrow\! \eta N$ vertex: 
\begin{equation} 
k + p = k^{\prime} + p^{\,\prime} \;,  
\end{equation} 
where $p$ is the four-momentum of the bound nucleon, whose spatial 
part is known in the literature as the missing momentum 
\begin{equation} 
{\bf p}_{m}\equiv {\bf p}^{\,\prime}-{\bf q} \;; \quad 
                 ({\bf q}\equiv{\bf k}-{\bf k}^{\prime}) \;. 
\end{equation} 
The dependence of the cross section on the missing momentum (See 
Section~\ref{sec:results}) results in the interaction becoming a 
probe of the nucleon momentum distribution, just as in most 
semi-inclusive processes, one of which is the $(e,e'p)$ reaction~\cite{AP01}.

\subsection{Cross section and polarization observables} 
\label{subsec:obs} 
 
Using the conventions of Bjorken and Drell \cite{MS84}, the differential
cross section for the scattering process depicted in Fig.~\ref{fig1} is
given by
\begin{equation} 
\left(\frac{d^5\sigma (s^{\prime},\varepsilon)} 
  {d{\bf k}^{\prime}d\Omega_{{\bf k}^{\prime}} 
   d\Omega_{{\bf p}^{\prime}}}\right)_{\rm lab} =  
   \frac{2 \pi}{2 E_\gamma} \mbox{ } 
   \frac{|{\bf k}^{\prime}|^2}{(2\pi)^3\,2E_{{\bf k}^{\prime}}}\mbox{ } 
   \frac{M_N|{\bf p}^{\prime}|}{(2\pi)^3}\mbox{ }  
   {\big|{\cal M}\big|}^2 \;, 
 \label{d5sigma} 
\end{equation} 
where $s^{\prime}$ is the spin of the emitted nucleon,  
$\varepsilon$ is the polarization of the incident photon,  
$M_N$ is the nucleon mass, and ${\cal M}$ is the scattering 
matrix element defined as 
\begin{equation} 
 \big|{\cal M}\big|^2 = \sum_m  
 \Big|  
  \overline{\cal U}({\bf p}^{\prime},s^{\prime})\mbox{ }  
  T(s,t)\mbox{ }{\cal U}_{\alpha,m}({\bf p}) 
 \Big|^2 \;. 
\label{Msquare} 
\end{equation}      
Note that ${\cal U}({\bf p}^{\prime},s^{\prime})$ is the free 
Dirac spinor for the knocked-out nucleon while  
${\cal U}_{\alpha,m}({\bf p})$ is the Fourier transform of  
the relativistic spinor for the bound nucleon ($\alpha$ 
stands for the collection of all quantum numbers of the 
single-particle orbital). The use of 
the impulse approximation is evident in this expression where we invoke the on-shell photoproduction operator $T(s,t)$.

By summing over the two spin components of the emitted nucleon and 
averaging over the transverse photon polarization in 
Eq.~(\ref{d5sigma}), we obtain an expression for the unpolarized 
differential cross section 
\begin{equation} 
\label{d5sigma0}
  \left(\frac{d^5\sigma} 
  {d{\bf k}^{\prime}d\Omega_{{\bf k}^{\prime}} 
   d\Omega_{{\bf p}^{\prime}}}\right)_{\rm lab} =  
   \frac{1}{2}\sum_{s^{\prime},\varepsilon} 
   \left(\frac{d^5\sigma (s^{\prime},\varepsilon)} 
   {d{\bf k}^{\prime}d\Omega_{{\bf k}^{\prime}} 
   d\Omega_{{\bf p}^{\prime}}}\right)_{\rm lab} \;. 
\end{equation} 
Nonetheless, our main interest is in the polarization observables as they 
are the true discriminators of subtle dynamics. Hence, the recoil 
nucleon polarization 
(${\cal P}$) is given by~\cite{wcc90,ndu91} 
\begin{equation}
\label{eq_pol}
{\cal P} = \sum_{\varepsilon}  
 \left( 
  \frac{{d^5\sigma}(\uparrow) - {d^5\sigma}(\downarrow)}  
       {{d^5\sigma}(\uparrow) + {d^5\sigma}(\downarrow)} 
 \right)_{\rm lab} \;, 
\end{equation} 
while the photon asymmetry ($\Sigma$) is provided through the 
expression~\cite{lwbt96,blmw98}  
\begin{equation}
\label{eq_asym}
 {\Sigma} = \sum_{s^{\prime}}  
  \left( 
   \frac{{d^5\sigma}(\perp) - {d^5\sigma}(\parallel)}  
        {{d^5\sigma}(\perp) + {d^5\sigma}(\parallel)} 
  \right)_{\rm lab} \;. 
\end{equation} 
Here the $\uparrow$ and $\downarrow$ represent the 
projection of the spin of the nucleon with respect to  
the normal to the scattering plane (${\bf k}\times{\bf k}^{\prime}$), 
while $\perp$ ($\parallel$) represents the out-of-plane (in-plane)  
polarization of the photon.

\subsection{Elementary ($\gamma p \rightarrow \eta N$) Amplitude} 
\label{subsec:elemapl}

We use the canonical model-independent parameterization for the 
elementary process $\gamma p \rightarrow \eta N$. This parameterization is constructed in  
terms of four Lorentz- and gauge-invariant  
amplitudes~\cite{cgln57,bt90,bt91} 
\begin{equation} 
\label{Tmatrix}
 T(\gamma p \rightarrow \eta N) \, = \,
 T(s,t) \, = \,
 \sum_{i=1}^{4} A_{i}(s,t) M_i\;, 
\end{equation}                              
where the invariant matrices have been defined as: 
\begin{subequations} 
\begin{eqnarray} 
 M_1 &=& - \gamma^5 \rlap/{\varepsilon} \rlap/k \;,   \\  
 M_2 &=& 2 \gamma^5 [(\varepsilon \cdot p) (k\cdot p^{\prime}) -  
         (\varepsilon\cdot p^{\prime}) (k\cdot p)]\;, \\  
 M_3 &=& \gamma^5[\rlap/{\varepsilon} ((k\cdot p) - \rlap/k 
         (\varepsilon \cdot p)], \; \\  
 M_4 &=& \gamma^5 [\rlap/{\varepsilon} ((k\cdot p^{\prime}) -  
         \rlap/k (\varepsilon \cdot p^{\prime})] \;. 
\label{M1234} 
\end{eqnarray} 
\end{subequations} 
One must stress here that although this parameterization is 
conventional, it is certainly not unique. Many other parameterizations, 
which are all 
equivalent for on-shell spinors, are possible~\cite{APSR99}. The 
problem is that in the quasifree process the bound 
nucleon is off its mass shell and these various parameterizations may no 
longer be equivalent off-shell.
While 
one can use a specific model for the $\eta$ photoproduction process 
which fixes the off-shell behavior of these amplitudes, this still does not 
solve the problem of medium modifications as the driving dynamics, such 
as the $S_{11}$ resonance, may change its properties in the nuclear 
medium. The problem can be addressed by modeling how 
the propagation in the nuclear medium affects the inherent 
dynamics. Such models, however, can be very different in their 
predictions  (See Section~\ref{intro}) which makes the quasifree process an ideal candidate to 
distinguish between them.

It is more convenient at this stage to transform the above parameterizations to a 
form that shows explicitly the Lorentz and parity structure of the amplitude as 
given by~\cite{PSB97,APSR99} 
\begin{equation} 
 T(\gamma p \rightarrow \eta N) =  
 F^{\alpha \beta}_{T} \sigma_{\alpha \beta} +  
 iF_P\gamma_5 +  
 F^{\alpha}_{A} \gamma _\alpha \gamma_5 \;, 
\end{equation} 
where the tensor, pseudoscalar, and axial-vector amplitudes are 
defined through 
\begin{subequations} 
\begin{eqnarray} 
 F^{\alpha \beta}_{T} &=& \frac {1}{2}  
  \varepsilon^{\mu\nu\alpha\beta}\varepsilon_\mu k_\nu A_1(s,t)\;, \\ 
 F_P &=&-2\,i\Big[(\varepsilon\cdot p) (k\cdot p^{\prime})  
        -(\varepsilon \cdot p^{\prime}) (k\cdot p)\Big]A_2(s,t)\;, \\ 
 F^{\alpha}_{A} &=& \Big[(\varepsilon \cdot p) k^\alpha - (k\cdot p)  
                 \varepsilon^\alpha\Big]A_3(s,t) +  
                 \Big[(\varepsilon \cdot p^{\prime})k^\alpha -  
                 (k\cdot p^{\prime})\varepsilon^\alpha\Big]A_4(s,t)\;. 
\end{eqnarray} 
\end{subequations}

While this parameterization is model independent, one still needs to 
determine the
four Lorentz and gauge invariant amplitudes $A_{i}(s,t)$. In principle 
one can just extract them experimentally or use some other formalism that 
calculates these amplitudes based on effective Lagrangian theory or other 
approaches. Here we use the effective Lagrangian approach of 
Benmerrouche, Zhang, and Mukhopadhyay (BZM)~\cite{benm92,mzb95,bmz95}. Such
 an approach is more fundamental and satisfies stringent theoretical
 constraints compared to other treatments such as Breit-Wigner-type
 parameterizations~\cite{Homma88,Hicks73} or coupled channel isobar
 models~\cite{bt90,lwbt96,TBK94}. The BZM treatment is specially
 distinguished by the limited number of free parameters (a maximum of
 eight) compared to other models.

The BZM aproach we adopt here, includes spin-1/2 [$S_{11}$(1535)] and spin-3/2 [$D_{13}$(1520)] 
resonances as well as nucleon Born terms and vector meson exchanges 
($\omega$ and $\rho$). All free parameters in the model were 
determined by fitting the experimental data for 
$p(\gamma,\eta)p$~\cite{krus95a}  
and $d(\gamma,\eta )pn$~\cite{krus95b}. The cross section is clearly
dominated by the $S_{11}$ resonance in this model. 
However, there are still important and necessary 
contributions, particularly in the angular distributions, from the 
$D_{13}$ resonance, nucleon born terms, and vector meson exchanges.

\subsection{Nuclear-Structure Model} 
\label{subsec:nsm}

We use in this work a relativistic 
mean-field approximation to the Walecka model~\cite{SW86} for all 
calculations of nuclear structure. In the case of nuclei with spherically 
symmetric potentials such as those studied here, the Dirac bound-state 
spinors can be 
classified with respect to a generalized angular momentum $\kappa$~\cite{sak73}. 
These states can be expressed in a two-component representation according to 
\begin{equation} 
 {\cal U}_{E \kappa m}({\bf x}) = \frac {1}{x}  
 \left[ \begin{array}{c} 
   g_{E \kappa}(x) {\cal Y}_{+\kappa \mbox{} m}(\hat{\bf x})  \\ 
  if_{E \kappa}(x) {\cal Y}_{-\kappa \mbox{} m}(\hat{\bf x}) 
 \end{array} \right], 
\end{equation} 
where the spin-angular functions are defined as: 
\begin{equation} 
 {\cal Y}_{\kappa\mbox{} m}(\hat{\bf x}) \equiv  
 \langle{\hat{\bf x}}|l{\scriptstyle\frac{1}{2}}jm>\;; \quad 
 j = |\kappa| - \frac {1}{2} \;; \quad 
 l =  
\begin{cases} 
\kappa\;,   & \text{if} \, \, \kappa>0; \\ 
            -1-\kappa\;,   & \text{if} \, \, \kappa<0.  
\end{cases} 
 \label{curlyy} 
\end{equation} 
The Fourier transform of the relativistic bound-state  
spinor is given by 
\begin{equation} 
  {\cal U}_{E\kappa m}({\bf p}) \equiv \int d{\bf x} \;  
      e^{-i{\bf p}\cdot{\bf x}}  \;  
      {\cal U}_{E\kappa m}({\bf x}) = 
      \frac{4\pi}{p} (-i)^{l} 
      \left[ 
      \begin{array}{c} 
       g_{E\kappa}(p) \\ 
       f_{E\kappa}(p) ({\bf \sigma}\cdot{\hat{\bf p}}) 
      \end{array} 
      \right] {\cal{Y}}_{+\kappa m}(\hat{\bf{p}}) \;, 
 \label{uofp} 
\end{equation}        
where we have written the Fourier transforms of the radial  
wave functions as 
\begin{subequations} 
\begin{eqnarray} 
   g_{E\kappa}(p) &=&  
    \int_{0}^{\infty} dx \,g_{E\kappa}(x)   
    \hat{\jmath}{\hbox{\lower 3pt\hbox{$_l$}}}(px) \;, \\ 
   f_{E\kappa}(p) &=& ({\rm sgn}\kappa)  
    \int_{0}^{\infty} dx \,f_{E\kappa}(x) 
    \hat{\jmath}{\hbox{\lower 3pt\hbox{$_{l'}$}}}(px) \;. 
\label{gfp} 
\end{eqnarray} 
\end{subequations} 
The Riccati-Bessel function is incorporated here in terms of the spherical 
Bessel function 
$\hat{\jmath}{\hbox{\lower 3pt\hbox{$_{l}$}}}(z)= 
zj{\hbox{\lower 3pt\hbox{$_{l}$}}}(z)$~\cite{Taylor72}. The 
$l'$ is the orbital angular momentum corresponding 
to $-\kappa$ as given in Eq.~(\ref{curlyy}).

\subsection{Closed-form Expression for the Photoproduction Amplitude} 
\label{subsec:closedform}

Having discussed some of the basic elements of our formalism, the next 
step is to calculate the square of the photoproduction amplitude  
[Eq.~(\ref{Msquare})]. In doing so we use the relativistic plane wave 
impulse approximation and incorporate no distortions for the emitted 
nucleon or $\eta$. Our rationale is that we concentrate on the 
polarization observables as they are true discriminators of the subtle 
dynamics. Fortunately, these observables, although very sensitive to 
the elementary amplitude dynamics, are strikingly insensitive to 
distortions as has been suggested by several nonrelativistic 
studies~\cite{lwb93,lwbt96,blmw98}. Consequently, it is 
straightforward to evaluate the knocked-out nucleon propagator using 
the Casimir ``trick'': 
\begin{equation} 
 S(p^{\prime}) \equiv \sum_{s^{\prime}} 
            {\cal U } ({\bf p}^{\prime},s^{\prime}) \, 
  \overline{{\cal U }}({\bf p}^{\prime},s^{\prime}) = 
  \frac{\rlap/p^{\prime}+M_{N}}{2M_{N}} \;; \quad 
  \left(p^{\prime\,0}\equiv E_{N}({\bf p}^{\prime})= 
   \sqrt{{\bf p}^{\prime\,2}+M_{N}^2}\right) \;. 
\label{sfree} 
\end{equation} 
Gardner and Piekarewicz~\cite{gp94,thesis,AP00} have shown that a 
similar procedure holds even for bound-state 
spinors where the ``bound-state propagator'' can be cast in the form 
\begin{eqnarray} 
  S_{\alpha}({\bf p})   
  &\equiv& \frac{1}{2j+1} \sum_{m}  
               {\cal U}_{\alpha,m}({\bf p}) \, 
      \overline{\cal U}_{\alpha,m}({\bf p}) \nonumber \\ 
  &=& \left(\frac{2\pi}{p^{2}}\right) 
      \left(  
       \begin{array}{cc} 
         g^{2}_{\alpha}(p) &  
        -g_{\alpha}(p)f_{\alpha}(p){\bf \sigma}\cdot{\hat{\bf p}} \\ 
        +g_{\alpha}(p)f_{\alpha}(p){\bf \sigma}\cdot{\hat{\bf p}} & 
        -f^{2}_{\alpha}(p)  
       \end{array} 
      \right) \nonumber \\ 
  &=& ({\rlap/{p}}_{\alpha} + M_{\alpha}) \;, \quad 
      \Big(\alpha=\{E,\kappa\}\Big) \;. 
 \label{salpha} 
\end{eqnarray} 
Note that we have defined the above mass-, energy-, and momentum-like 
quantities as 
\begin{subequations} 
\begin{eqnarray} 
  M_{\alpha} &=& \left(\frac{\pi}{p^{2}}\right) 
                  \Big[g_{\alpha}^{2}(p) - 
                       f_{\alpha}^{2}(p)\Big] \;, \\ 
  E_{\alpha} &=& \left(\frac{\pi}{p^{2}}\right) 
                  \Big[g_{\alpha}^{2}(p) + 
                       f_{\alpha}^{2}(p)\Big] \;, 
 \label{epm} \\ 
  {\bf p}_{\alpha} &=& \left(\frac{\pi}{p^{2}}\right) 
                   \Big[2 g_{\alpha}(p)  
                          f_{\alpha}(p)\hat{\bf p}  
                   \Big]  \;, 
\end{eqnarray} 
\end{subequations} 
which satisfy the ``on-shell relation'' 
\begin{equation} 
  p_{\alpha}^{2}=E_{\alpha}^{2}-{\bf p}_{\alpha}^{2} 
                =M_{\alpha}^{2} \;. 
 \label{onshell} 
\end{equation} 
This algebraic trick results in an enormous simplification in the 
formalism due to the similarity in structure between the free and 
bound propagators [Eqs.~(\ref{sfree}) and (\ref{salpha})]. The calculation of the square of the photoproduction 
amplitude boils down to an evaluation of traces of $\gamma$-matrices. 
This simplification would have not been possible had we incorporated 
distortion effects for the emitted nucleon. 
 
In evaluating the ensuing traces of $\gamma$ matrices, we are greatly aided 
by the use of {\it FeynCalc} package~\cite{MBD91} with {\it 
Mathematica}. The tedious work of calculating, analytically, the tens 
of traces of $\gamma$ matrices can be borne by the computer and 
eventually we arrive elegantly at transparent analytical results for 
all observables. The results can then be fed into a FORTRAN code to obtain the final numerical values for 
all observables in the problem.

\section{Results and Discussion} 
\label{sec:results} 
 
There are many ingredients, mostly related to the nature of the model used
for the elementary process, 
that go into the calculation of the observables for
quasifree $\eta$  photoproduction. 

In our treatment, the basic ingredients are
the Born and vector meson terms, the
$S_{11}$ and $D_{13}$ resonances, 
as well as the coupling constants and the choice of
pseudoscalar or pseudovector coupling at the meson-nucleon vertex. In the
subsequent sections we investigate the sensitivity of the observables to 
variations in the elementary amplitude,  medium
modifications to the masses of the $S_{11}$ and $D_{13}$ resonances as
well as to different nuclear targets. All calculations are done
not far from threshold at an incident photon laboratory 
kinetic energy of 750 MeV.

In Ref. \cite{lwbt96}
a nonrelativistic distorted wave model of quasifree $\eta$
photoproduction was given. In the previous section we presented a
fully relativistic formalism, but using the plane wave approximation
for the nucleon and $\eta$. While distortions are very difficult to
incorporate in an analytical relativistic treatment such as ours, they
do not appear to have any significance apart from quenching the cross
section in such a manner that does not affect polarization
observables. Indeed, the spin observables although very sensitive to the
elementary amplitude dynamics, are insensitive to distortions at
least as far as the $\eta$ photoproduction process is concerned~\cite{lwb93,lwbt96,blmw98}.
We will therefore pay special attention to these observables in all subsequent sections.

\subsection{Sensitivity to variations in the elementary amplitude}
\label{sec:elementary_amplitude}

It is clear from Eqs. (\ref{d5sigma}) and
(\ref{d5sigma0})-(\ref{eq_asym}) that the unpolarized differential
cross section, the recoil nucleon
polarization and the photon asymmetry are determined by the scattering
matrix element ${\cal M}$. Eq. (\ref{Msquare}) in turn shows that one
of the primary ingredients in the calculation of ${\cal M}$ is the
interaction matrix $T(s,t)$. From Eq. (\ref{Tmatrix}) we see that it
is determined by the four invariant amplitudes $A_{i} (s,t)$. These
amplitudes contain the dynamics of the reaction. In this work the
effective Lagrangian model of Benmerrouche, Zhang and Mukhopadyay
\cite{benm92,mzb95,bmz95} is used to determine the dynamics where the
elementary process depends on four key components: (i) the nucleon Born terms, (ii)
the vector mesons ($\omega$ and $\rho$) exchange, (iii) the $S_{11}$
resonance and (iv) the $D_{13}$ resonance. The contributions from
other vector mesons such as the $\phi$ or other heavier mesons are
shown to be negligible as a consequence of several reasons including the Okubo-Zweig-Iizuka
suppression, the largeness of masses, and the meager coupling to the
photoproduction channel~\cite{bmz95}. Furthermore, 
the contributions from other resonances are
insignificant primarily due to their larger masses or small coupling
to the $\eta N$ channel in the energy regime of interest (near threshold)~\cite{bmz95}.
In this section we
investigate how each of these four components influence the measured
observables. All results in this section were obtained for proton
knockout from the $1p^{3/2}$ orbital of $^{12}$C, apart from the results in 
Fig.~\ref{calculational_types_theta_eta_N} which refer to neutron knockout from the
same orbital. All calculations were done for an incident photon
laboratory kinetic energy, $E_{\gamma} \, = \, 750$ MeV, while the missing momentum
was fixed at $|\vec{p}_{m} \, | \, = \, 100$ MeV. This value of $|\vec{p}_{m} \, |$
is near the peak of the momentum distribution of the bound nucleon, i.e., where 
the cross section is maximized. Except for Fig.~\ref{etaNN_vertex}, all calculations 
employed pseudoscalar coupling at the $\eta NN$ vertex.

\begin{figure}
\includegraphics[width=8.5cm,angle=0]{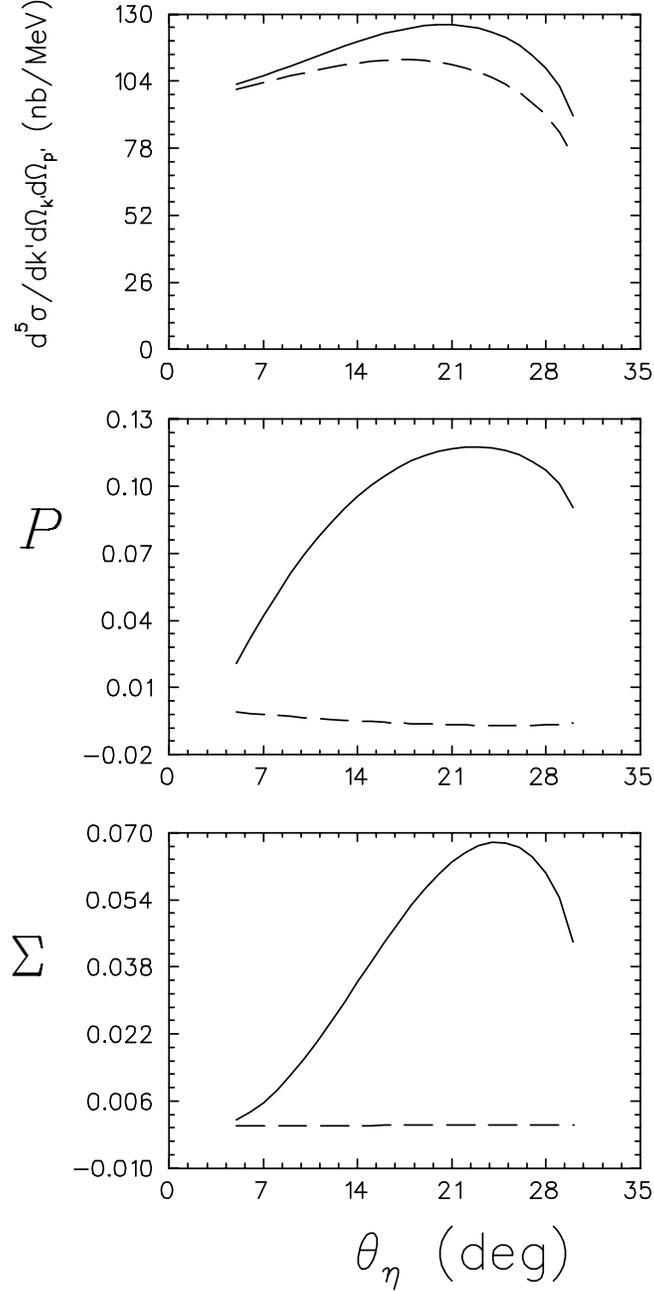}
 \caption{\label{calculational_types_theta_eta1} Unpolarized
 differential cross section $(d^{5}\sigma/dk'd\Omega_{k'} d
 \Omega_{p'})$, recoil nucleon polarization $({{\cal P}})$ and photon
 asymmetry $(\Sigma)$ as a function of the $\eta$ meson scattering angle
 $\theta_{\eta}$ for proton knockout from the $1p^{3/2}$ orbital of
 $^{12}$C.  The solid line represents the calculation using the Born
 and vector meson terms together with both the $S_{11}$ and $D_{13}$
 resonances.  The dashed line represents the calculations employing
 only the $S_{11}$ resonance. The incident photon laboratory kinetic
 energy is $E_{\gamma} \, = \, 750$ MeV and the missing momentum is
 fixed at $|\vec{p}_{m} \, | \, = \, 100$ MeV.  }
\end{figure}

In Fig.~\ref{calculational_types_theta_eta1} we show 
calculations of the unpolarized
differential cross section $(d^{5} \sigma / dk' d \Omega_{k'} d \Omega_{p'})$, 
the recoil nucleon polarization $({\cal P})$ and the photon
asymmetry $(\Sigma)$ for proton
knockout from the $1p^{3/2}$ orbital of $^{12}$C
as a function of 
the $\eta$ meson scattering angle, $\theta_{\eta}$.
In this figure the solid line represents the full calculation where all four 
components of the effective Lagrangian model are used in the evaluation of the 
invariant amplitudes $A_{i}(s,t)$. The dashed line represents the
calculation using only the $S_{11}$ resonance. 
Fig.~\ref{calculational_types_theta_eta1} clearly displays that the $S_{11}$
resonance gives the dominant contribution to the unpolarized quasifree
differential cross section just as in the elementary reaction. This is in contrast to coherent $\eta$ photoproduction where the 
contribution of this resonance is strongly suppressed due to the
filtering of the isoscalar component and the spin-flip
nature of the $S_{11}$ contribution~\cite{PSB97,APSR99,PLM98}. 
Despite the $S_{11}$ dominance of the differential cross section, the
nucleon polarization and photon asymmetry yield only small
contributions from the $S_{11}$ resonance. This illustrates
beautifully how the polarization observables are the true 
discriminators of subtle dynamics. While the differential cross
section is largely blind to all but the $S_{11}$ contribution, the
polarization observables show the intricacies of the dynamics in full
color.

\begin{figure}
\includegraphics[width=12cm,angle=0]{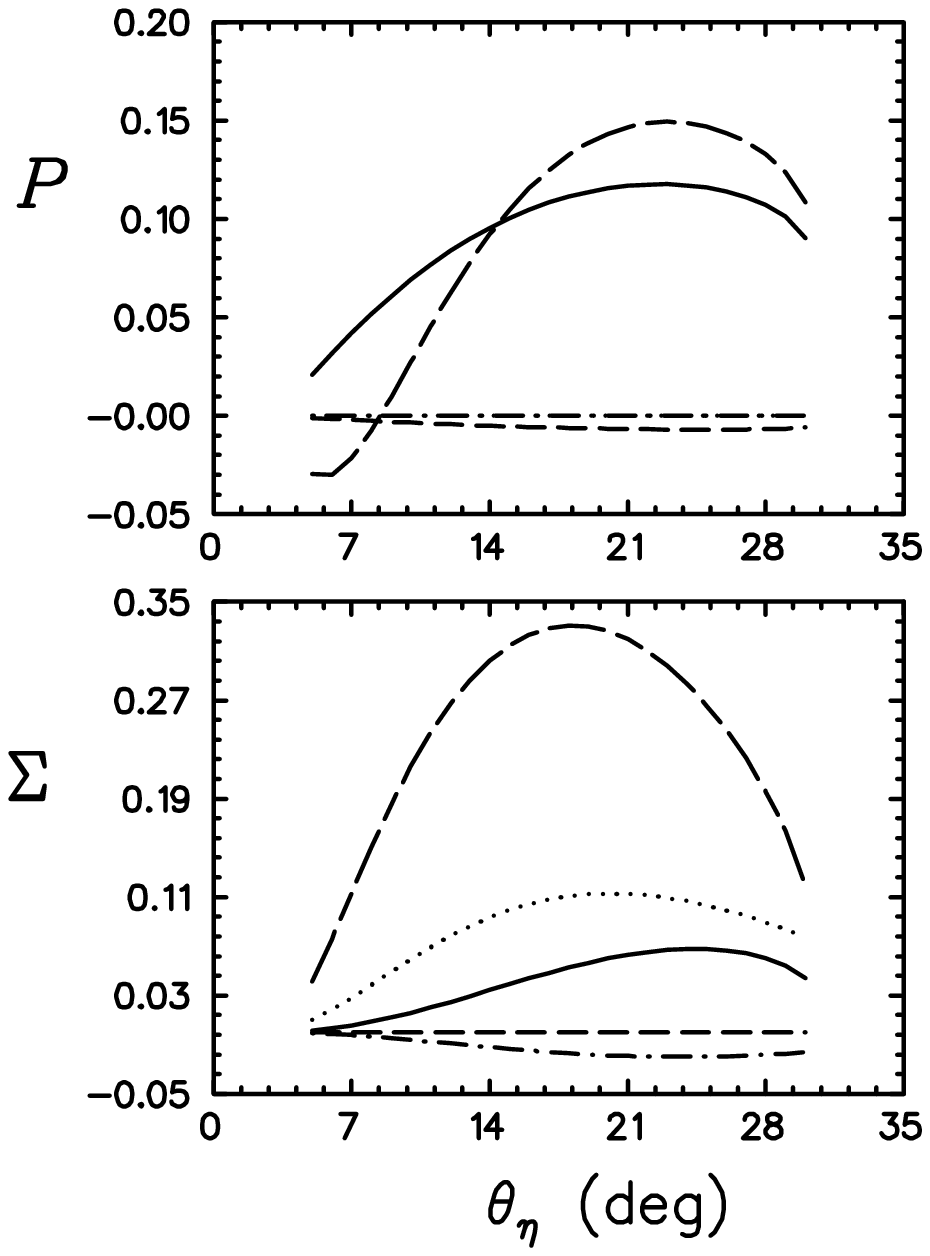}
 \caption{\label{calculational_types_theta_eta} The recoil nucleon
 polarization $({\cal P})$ and the photon asymmetry ${(\Sigma)}$ as a
 function of the $\eta$ meson scattering angle $(\theta_{\eta})$ for
 proton knockout from the 1$p^{3/2}$ orbital of $^{12}$C.  The
 incident photon energy is $E_{\gamma} \, = \, 750$ MeV and the
 missing momentum is fixed at $ | \vec{p}_{m}| \, = \, 100$ MeV.  The
 solid line represents the full calculation employing both the
 $S_{11}$ and $D_{13}$ resonances together with the Born and vector meson
 terms. The dashed and long-dash short-dash lines represent employing
 only the $S_{11}$ and $D_{13}$ resonances, respectively.  The
 dash-dot and dotted lines correspond to the calculation employing
 only the Born terms and vector meson terms, respectively.  }
\end{figure}

\begin{figure}
\includegraphics[width=12cm,angle=0]{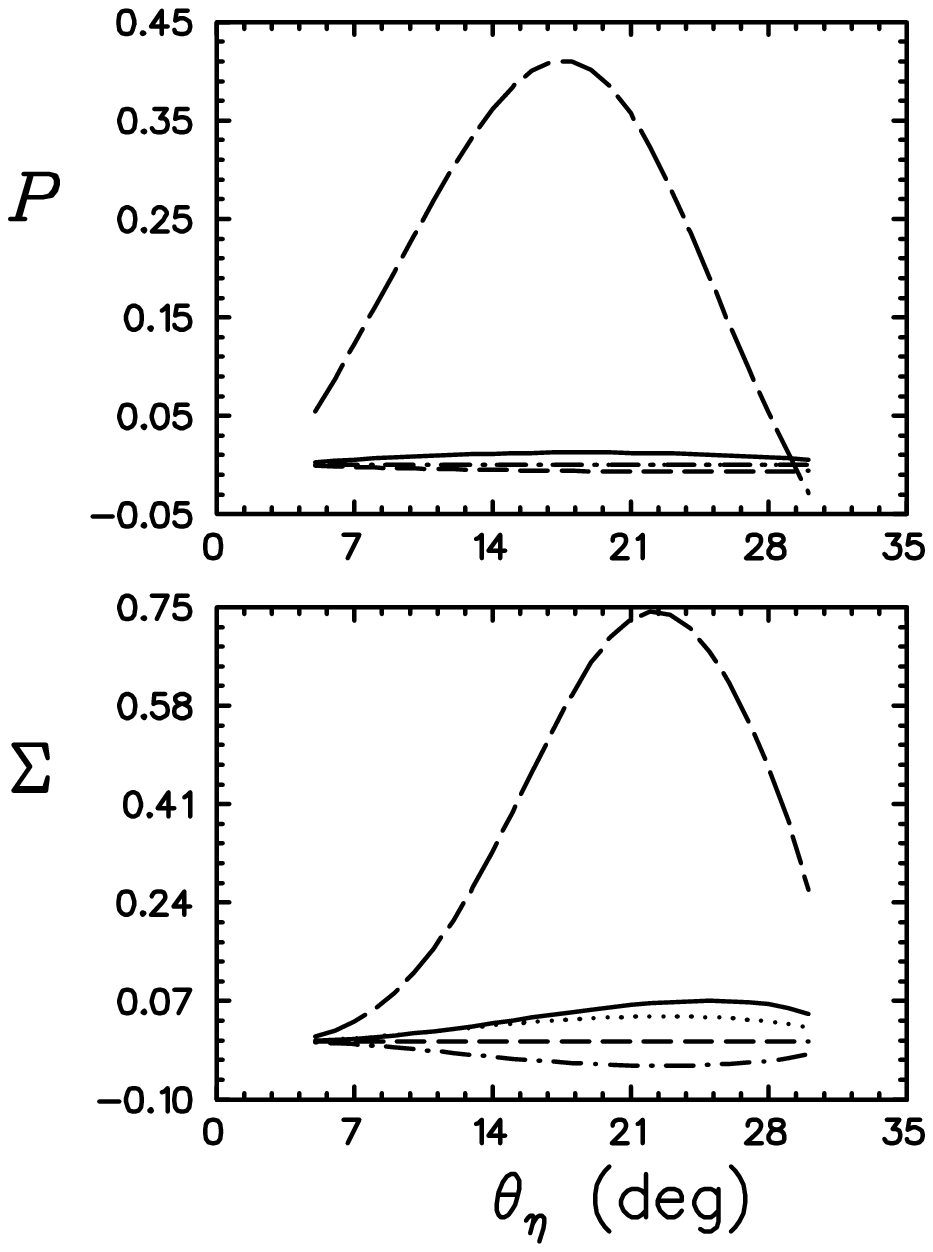}
 \caption{\label{calculational_types_theta_eta_N} The recoil nucleon
 polarization $({\cal P})$ and the photon asymmetry ${(\Sigma)}$ as a
 function of the $\eta$ meson scattering angle $(\theta_{\eta})$ for
 neutron knockout from the 1$p^{3/2}$ orbital of $^{12}$C.  The
 incident photon energy is $E_{\gamma} \, = \, 750$ MeV and the
 missing momentum is fixed at $ | \vec{p}_{m}| \, = \, 100$ MeV.  The
 solid line represents the full calculation employing both the
 $S_{11}$ and $D_{13}$ resonances together with the Born and vector meson
 terms. The dashed and long-dash short-dash lines represent employing
 only the $S_{11}$ and $D_{13}$ resonances, respectively.  The
 dash-dot and dotted lines correspond to the calculation employing
 only the Born terms and vector meson terms, respectively.}
\end{figure}

\begin{figure}
\includegraphics[height=12cm,angle=0]{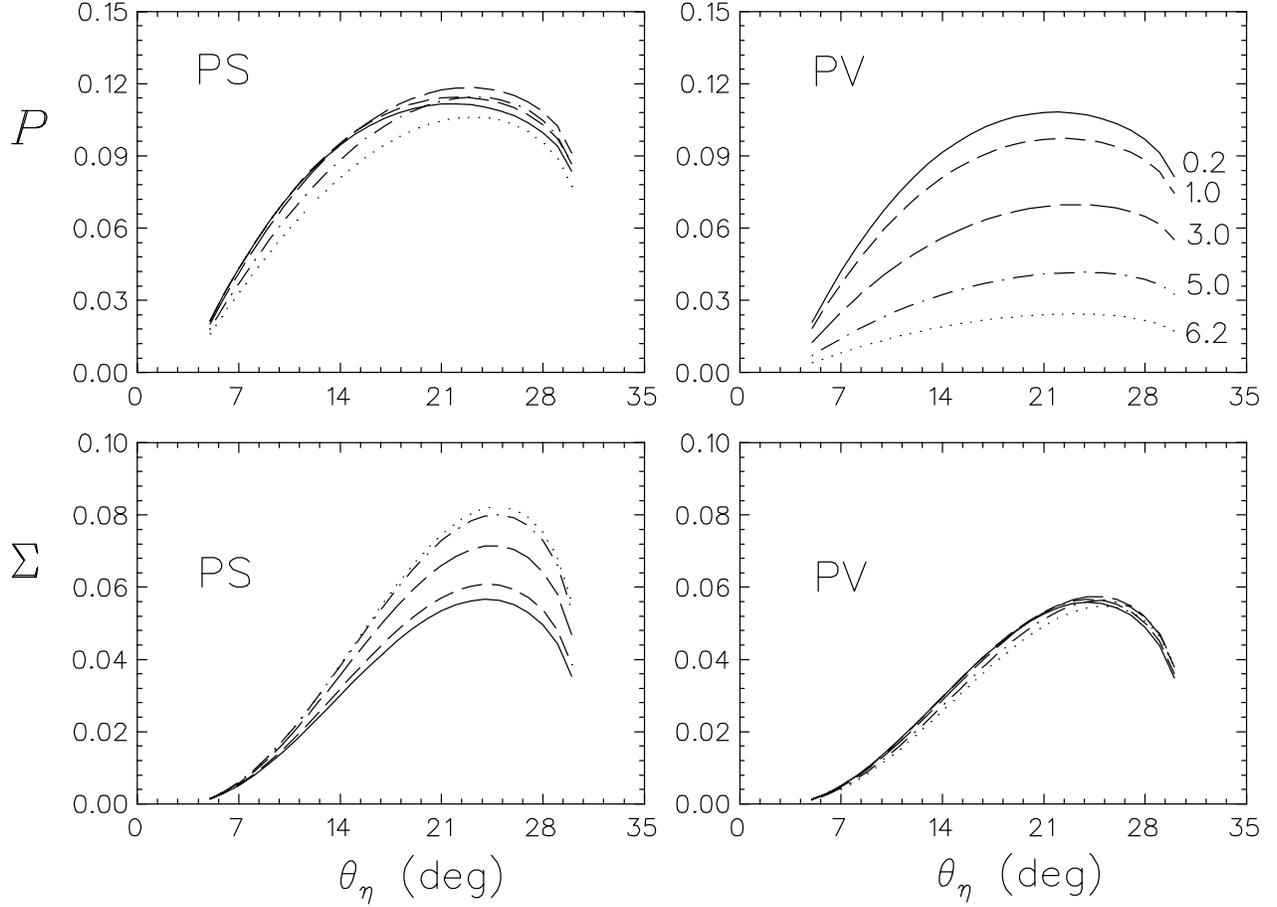}
 \caption{\label{etaNN_vertex} Variation of the recoil nucleon
 polarization $({\cal P})$ and the photon asymmetry $(\Sigma)$ with
 respect to a change in the value of the $\eta$ meson coupling
 constant $(g_{\eta})$ where $0.2 \, \le \, g_{\eta} \, \le \,
 6.2$. The graphs on the left-hand-side (right-hand-side) are for PS
 (PV) coupling at the $\eta NN$ vertex. The values of the coupling
 constant chosen in the range specified above are shown on the graph
 of the polarization for pseudovector coupling.  The calculations
 shown are for proton knockout from the 1$p^{3/2}$ orbital of $^{12}$C
 for an incident photon energy of 750 MeV and fixed missing momentum,
 $ | \vec{p}_{m}| \, = \, 100$ MeV. The calculation employed both the
 $S_{11}$ and $D_{13}$ resonances together with the Born and vector meson
 terms.  }
\end{figure}

In Fig.~\ref{calculational_types_theta_eta} the solid line represents
the full calculation employing both the $S_{11}$ and $D_{13}$
resonances together with the Born and vector meson terms.  The dashed and
long-dash short-dash lines represent employing only the $S_{11}$ and
$D_{13}$ resonances, respectively. The dash-dot and dotted lines
correspond to the calculation employing only the Born and vector meson
terms, respectively. In contrast to the unpolarized differential cross
section, the recoil nucleon polarization and the photon asymmetry are
not dominated by the $S_{11}$ but the $D_{13}$ resonance makes
the largest contribution.  The $S_{11}$ resonance, as well as the Born
and vector meson terms, make significant contributions to the spin
observables but through interference terms of the large $S_{11}$
amplitude and the small ones coming from the other background components in
the $\eta$ photoproduction process. It is easy to note that the full
calculation is significantly reduced with respect to the $D_{13}$
calculation for the photon asymmetry due to these interference effects.
These effects highlights the relevance of the polarization
observables in studying the background processes in the $\eta$
photoproduction process.

In Fig.~\ref{calculational_types_theta_eta_N} we show the polarization
and asymmetry for neutron knockout. As for the case of proton
knockout, all components of the elementary amplitude contributes
significantly to the polarization observables specially through the
interference terms which are stronger for the case of the neutron compared
to that of the proton. Moreover, the $D_{13}$ resonance makes the largest
contribution for these observables.
The branching ratios of the $S_{11}$ and $D_{13}$ resonances are
$50\%$ and $0.1\%$, respectively \cite{bmz95}. The sensitivity of the
polarization observables to the $D_{13}$ resonance is therefore
striking in light of its small branching ratio. However, this
sensitivity is very advantageous since the polarization observables
therefore provide a unique opportunity to study the properties of this spin 3/2
resonance such as its off-shell effects which are believed to be very
important~\cite{bmz95}. Note that in both 
Figs.~\ref{calculational_types_theta_eta} and 
\ref{calculational_types_theta_eta_N} the calculation employing only the
vector meson exchange, results in the polarization being identically
zero, hence no dotted line appears in either of these two figures for the
polarization. The reason for this is that the "potential" resulting from
only vector meson exchange is spin independent.

With respect to the Born terms, there are two uncertainties concerning
the $\eta NN$ vertex: (i) the magnitude of the coupling constant and
(ii) the type of coupling, i.e., pseudoscalar (PS) or pseudovector
(PV) coupling. While for the $\pi NN$ vertex there are convincing
reasons, in terms of the low energy theorem and the approximate chiral
symmetry of the SU(2) $\times$ SU(2) group, to oblige us to favor the
PV over the PS coupling~\cite{DeBaenst70,BKM92}, such considerations
are not necessarily valid for the $\eta NN$ vertex as a consequence of
the largeness of the $\eta$ mass and the significant breaking of the
chiral SU(3) $\times$ SU(3) group~\cite{GMOR68,MG84}.  In
Ref. \cite{bmz95} it is shown that an acceptable range for the $\eta
NN$ coupling constant $(g_{\eta})$ is:
\begin{eqnarray}
\label{g_eta_range}
 0.2 \, \le & g_{\eta} & \le \, 6.2.
\end{eqnarray}
In Fig.~\ref{etaNN_vertex} we show calculations of the recoil nucleon
polarization and the photon asymmetry for pseudoscalar and
pseudovector coupling when the value of $g_{\eta}$ is varied in the
range specified in Eq. (\ref{g_eta_range}).  The graphs on the
left-hand-side (right-hand-side) are for PS (PV) coupling at the $\eta
NN$ vertex. The values of the coupling constant chosen in the range
specified above are shown on the graph of the polarization for
pseudovector coupling.  For the case of pseudoscalar (pseudovector)
coupling, the polarization (asymmetry) is largely insensitive to
variations in the value of $g_{\eta}$. When pseudovector coupling is
employed, the polarization displays significant variations over the
entire angular range. In fact, it systematically decreases with an
increase in the value of $g_{\eta}$. Note that the asymmetry with
pseudoscalar coupling has exactly the opposite behavior in that it
decreases with a decrease in the value of $g_{\eta}$. In general the
pseudovector coupling decreases the polarization for a fixed value of
$g_{\eta}$. This decrease is more drastic the larger the value of
$g_{\eta}$ as can be seen by comparing the dotted line for the
polarization for pseudoscalar and pseudovector coupling. The
pseudovector coupling also tends to decrease the value of the photon
asymmetry. The nonrelativistic analysis of Ref. \cite{lwbt96} seems to
indicate that the data favors a pseudoscalar coupling at the $\eta NN$
vertex. This is in contrast to the analysis of the elementary process
in Ref. \cite{bmz95} where no conclusive evidence could be found for
either type of coupling. However, caution must be exercised when
comparing the two analyses, since the underlying dynamics are very
different. Our results indicate that the magnitude of $g_{\eta}$ also
plays a role in investigating this ambiguity. To emphasize:
the polarization is sensitive to the difference in the type of
coupling when $g_{\eta}$ is of the order of 3.0, 5.0 or 6.2.

\subsection{Sensitivity to medium modifications of the resonance masses}
\label{sec:resonance_medium_effects}

In this section we investigate the sensitivity of the nucleon polarization and
photon asymmetry to medium modifications of the masses of the $S_{11}$ and $D_{13}$ 
resonances. Since in the quasifree process the resonances propagate in
the nuclear medium as opposed to free space, one would expect changes
to their physical properties such as effective masses. Now we
examine the sensitivity of our observables to any possible medium
modifications to these masses.

\begin{figure}
\includegraphics[width=14cm,angle=0]{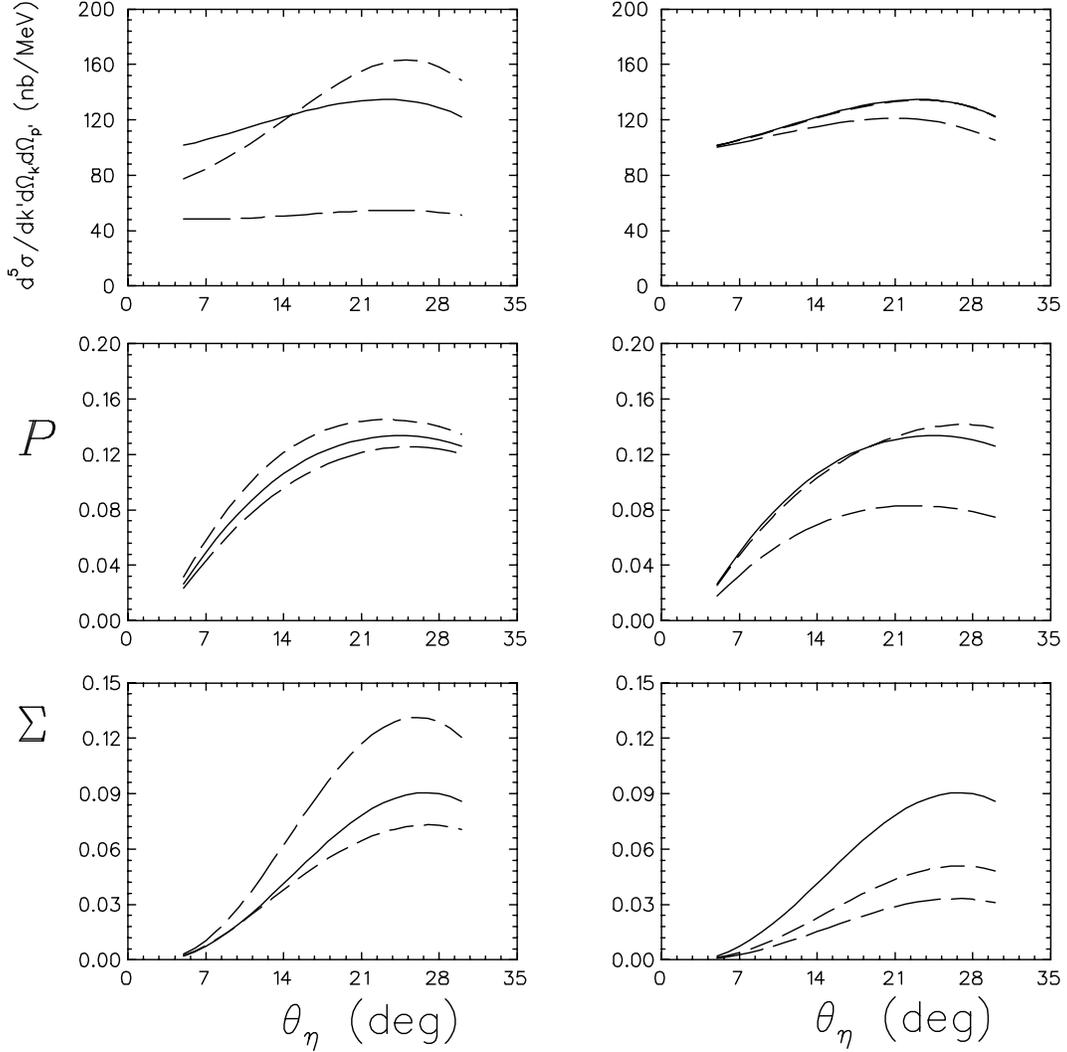}
 \caption{\label{resonance_medium_effects} Effect on the unpolarized
 differential cross section $(d^{5} \sigma / dk' d \Omega_{k'} d
 \Omega_{p'})$, recoil nucleon polarization $({\cal P})$ and photon
 asymmetry $(\Sigma)$ when the masses of the $D_{13}$ and $S_{11}$
 resonances are increased (long-dash short-dash line) or decreased
 (dashed line) by 3\%. The solid line corresponds to the free mass
 value for the two resonances.  The results shown are for proton
 knockout from the $1p^{3/2}$ orbital of $^{12}$C employing both
 resonances together with the Born and vector meson terms. The incident
 photon energy is $E_{\gamma} \, = \, 750$ MeV and the missing
 momentum is fixed at $ | \vec{p}_{m} \, | \, = \, 100$ MeV. 
The graphs on the left-hand-side
 correspond to a variation in the mass of the $S_{11}$ resonance,
 whilst keeping the mass of the $D_{13}$ resonance fixed. The graphs
 on the right-hand-side correspond to a variation in only the mass of
 the $D_{13}$ resonance.  }
\end{figure}

We consider proton knockout from the 1$p^{3/2}$ orbital of $^{12}$C
with an incident photon energy of $E_{\gamma} \, = \, 750$ MeV and a
missing momentum of $|\vec{p}_{m} \, | \, = \, 100$ MeV. For this
calculation we used the Born and vector meson terms together with both
resonances.  In Fig.~\ref{resonance_medium_effects} the solid line
corresponds to the free mass values, the dashed line a decrease of 3\%
and the long-dash short-dash line an increase of 3\% in the masses of
the $S_{11}$ and $D_{13}$ resonances. The graphs on the left-hand-side
of Fig.~\ref{resonance_medium_effects} show the sensitivity of the
polarization and the asymmetry to a change in the mass of the $S_{11}$
resonance, whilst keeping the mass of the $D_{13}$ resonance
fixed. For the graphs on the right-hand-side, only the mass of the
$D_{13}$ resonance was varied.

The nucleon polarization shows little sensitivity over the entire
angular range when only the mass of the $S_{11}$ resonance is
varied. 
The polarization is not sensitive to a decrease in the mass of the
$D_{13}$ resonance. An increase in the mass of $D_{13}$ leads to a
significant reduction relative to the free mass calculation for the
polarization. 
The
photon asymmetry does exhibit some sensitivity to a change in the mass
of $S_{11}$.  This observable is more sensitive to variations in
the mass of $D_{13}$, in particular an increase. This confirms the
findings of Ref.~\cite{lwbt96} that the photon asymmetry is a very
useful observable to look for medium modifications for the resonances.
Our results now indicate that in addition, the polarization is also
sensitive to medium modifications to the mass of the $D_{13}$
resonance.

\subsection{Sensitivity to the nuclear target}
\label{sec:nuclear_target}

\begin{figure}
\includegraphics[width=13cm,angle=0]{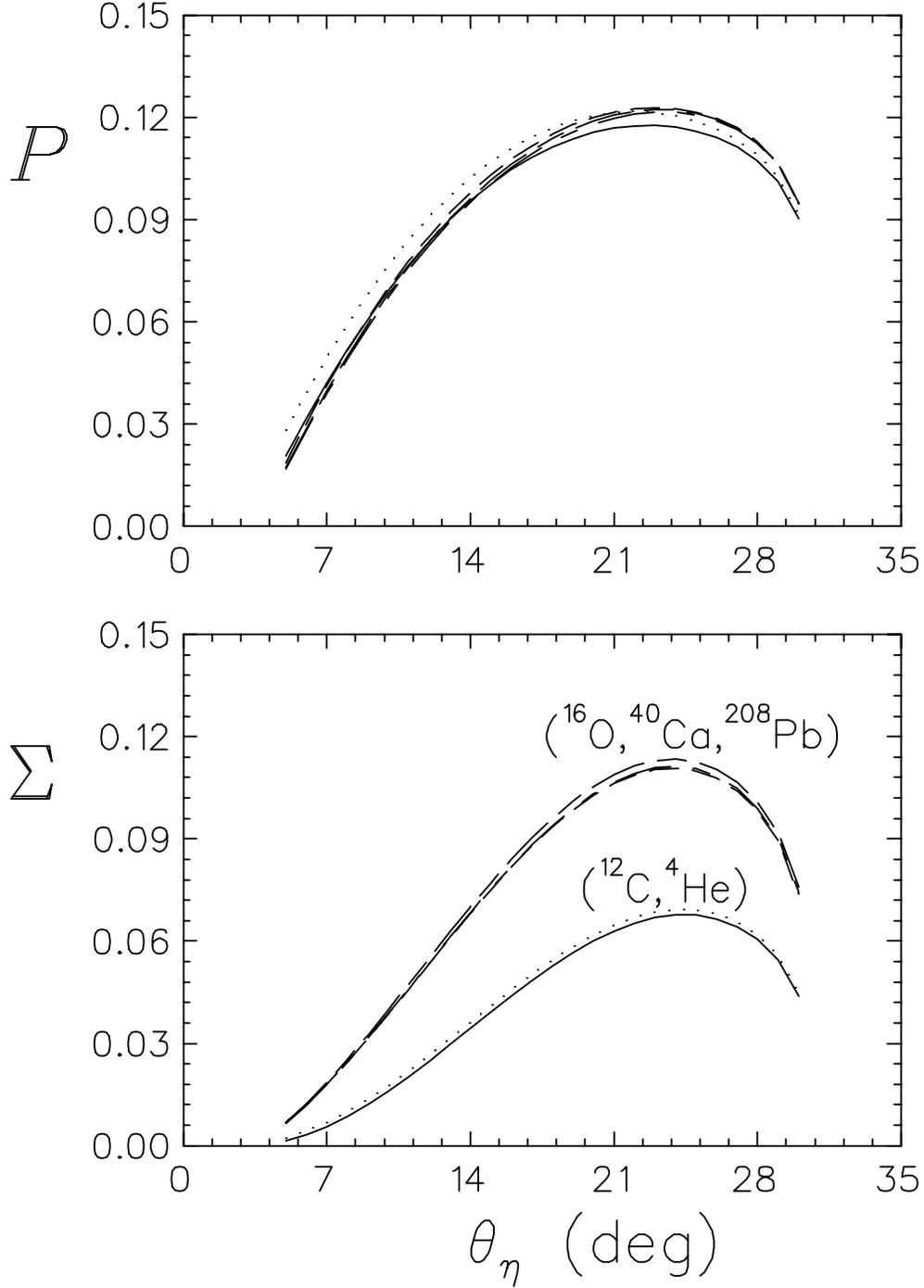}
 \caption{\label{nuclear_target_effects} Recoil nucleon polarization
 $({\cal P})$ and photon asymmetry $(\Sigma)$ for a variety of nuclear
 targets as a function of $\theta_{\eta}$. The results shown are for
 the knockout of valence protons from the particular nucleus and
 employed both the $S_{11}$ and $D_{13}$ resonances together with the
 Born and vector meson terms.  The incident photon energy is $E_{\gamma} \,
 = \, 750$ MeV and the missing momentum is $| \vec{p}_{m} \, | \, = \,
 100$ MeV. The curves corresponding to a particular group of nuclei
 are shown on the graph of the asymmetry.  }
\end{figure}

In Fig.~\ref{nuclear_target_effects} we show results for the nucleon
polarization and 
the photon asymmetry as a function of 
$\theta_{\eta}$ for a variety of nuclear targets. We considered valence proton
knockout from the 1$s^{1/2}$ orbital of $^{4}$He, the 1$p^{3/2}$ orbital of
$^{12}$C, the 1$p^{1/2}$ orbital of $^{16}$O, the 1$d^{3/2}$ orbital of
$^{40}$Ca and the 3$s^{1/2}$ orbital of $^{208}$Pb. The incident photon
energy was taken to be $E_{\gamma} \, = \, 750$ MeV and the missing
momentum $| \vec{p}_{m} \, | \, = \, 100$ MeV. 
The calculations employed the Born and
vector meson terms together with both the $S_{11}$ and $D_{13}$ resonances, 
as well as pseudoscalar coupling. 
The calculations clearly
indicate that the nucleon polarization is practically target independent.
This finding is in agreement with that of
Ref. \cite{AP00} (although for the kaon quasifree process).
In the relativistic case the photon asymmetry does indeed exhibit some
dependence on the 
nuclear target although the sensitivity is rather small. The
calculation for $^{12}$C and $^{4}$He coincide, 
while the photon  asymmetry for 
$^{16}$O, $^{40}$Ca and $^{208}$Pb are practically
indistinguishable. The apparent independence of the polarization
observables to nuclear target effects appears to be a feature
shared by various meson photoproduction processes~\cite{AP00}.

\subsection{Unpolarized differential cross section as a measure of the 
momentum distribution of the wavefunction}
\label{sec:xsection_momentum_distribution}

\begin{figure}
\includegraphics[width=13cm,angle=0]{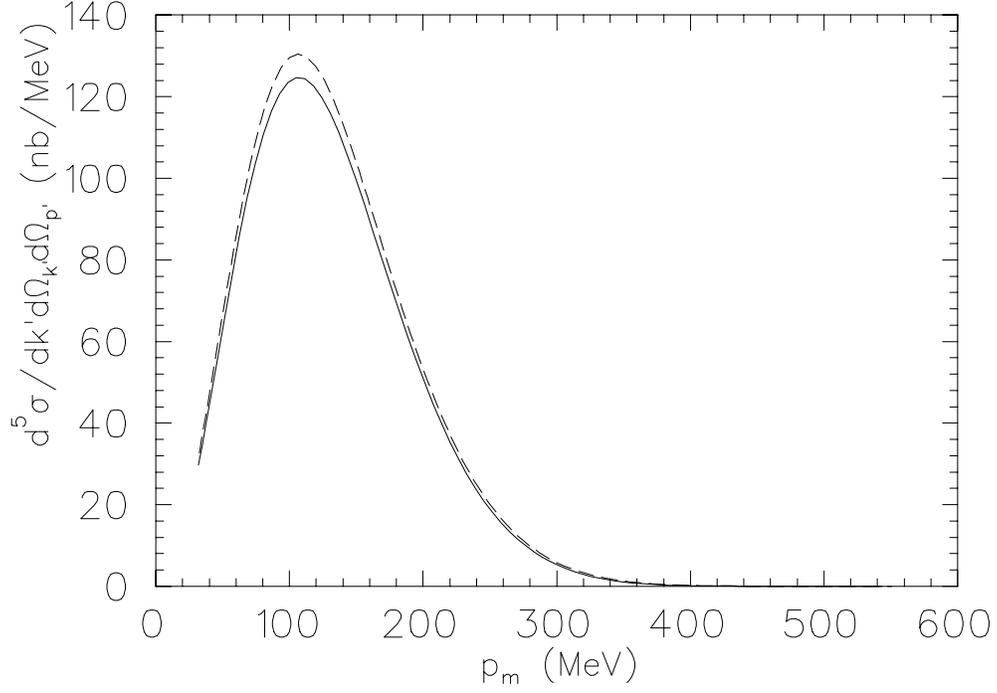}
 \caption{\label{sigma0} The unpolarized differential cross section (solid line) as
 a function of the missing momentum for proton knockout from the
 1$p^{3/2}$ orbital of $^{12}$C. The incident photon energy is
 $E_{\gamma} \, = \, 750$ MeV and the momentum transfer is fixed at
 $|\vec{q} \, | \, = \, 400$ MeV. The dashed lined represents the
 parameter $E_{\alpha}$ (up to an arbitrary scale) which is
 proportional to the momentum distribution of the bound proton
 wavefunction.}
\end{figure}

The momentum distribution of the bound nucleon is customarily measured
by doing electron scattering~\cite{AP01}. In this section we
illustrate the remarkable similarity between the cross section and the
momentum distribution of the bound nucleon for $\eta$ photoproduction.
In Fig.~\ref{sigma0} we show the unpolarized cross section (solid
line) as a function of the missing momentum for proton knockout from
the 1$p^{3/2}$ orbital of $^{12}$C. The incident photon energy was
taken to be $E_{\gamma} \, = \, 750$ MeV and the momentum transfer is
fixed at $| \vec{q} \, | \, = \, 400$ MeV. For this calculation we
have used the Born and vector meson terms together with both the
$S_{11}$ and $D_{13}$ resonances.  The dashed line represents the
parameter $E_{\alpha}$ (up to an arbitrary scale) which is
proportional to the momentum distribution of the bound proton
wavefunction (see Eq. \ref{epm}). The similarity between the momentum
distribution of the bound proton wavefunction and the cross section is
undeniable. Beyond 300 MeV the cross section quickly tends to zero. A
similar result was obtained in Ref. \cite{AP00} for kaon
photoproduction.

\section{Conclusions} 
\label{sec:concl} 
 
In this paper we have studied quasifree $\eta$ photoproduction via the
calculation of the differential cross section, the recoil nucleon polarization and the 
photon asymmetry using a relativistic plane wave impulse approximation formalism. 
The emphasis was on the polarization observables since they are
very sensitive to the underlying dynamics but largely insensitive to
distortion and nuclear target effects. The use of a plane wave formalism greatly
simplifies the calculation of the transition matrix element. Gardner
and Piekarewicz showed in Ref. \cite{gp94} that by introducing a
bound-state propagator one can still write $|{\cal M}|^{2}$ in terms
of traces over Dirac matrices. This not only allows one to use
Feynman's trace techniques even for quasifree scattering, but
additionally results in analytical expressions for the spin
observables. The boundstate wavefunction of
the bound nucleon was calculated within a relativistic mean-field
approximation to the Walecka model. For the elementary process we used the
effective Lagrangian approach of Benmerrouche, Zhang, and
Mukhopadhyay~\cite{benm92,mzb95,bmz95}.

We investigated the sensitivity of the various observables to the
elementary amplitude, medium modifications to the masses of the
$S_{11}$ and $D_{13}$ resonances as well as nuclear target
effects. Our results indicate that the nucleon polarization is
practically target independent, whereas the asymmetry exhibits some
small sensitivity. The polarization observables are very sensitive to
the elementary amplitude. We find that, in contrast to coherent $\eta$
photoproduction, the $S_{11}$ resonance completely dominates the
unpolarized cross section. However, the two spin observables are
dominated by the background processes in the elementary amplitude and
specially sensitive to the $D_{13}$ resonance contribution.
As a consequence, the polarization and asymmetry are
considerably sensitive to variations in the mass of the $D_{13}$
resonance. Indeed, a variation in the mass of this resonance leads to
significant effects in the polarization and asymmetry. This finding
agrees with the nonrelativistic analysis of Ref.  \cite{lwbt96}. The
polarization and asymmetry are useful tools to study the two ambiguities at
the $\eta NN$ vertex. The sensitivity of these observables to the
magnitude of the coupling constant depends to a large extent on the
type of coupling. The polarization (asymmetry) is insensitive to the
magnitude of the coupling constant for pseudoscalar (pseudovector)
coupling. However, the polarization (asymmetry) does indeed exhibit a
sensitivity to the magnitude of the coupling constant for pseudovector
(pseudoscalar) coupling.

The $\eta$ photoproduction process shares many features with kaon
photoproduction when viewed within a relativistic framework. There are
several differences when compared to the findings of the
nonrelativistic calculations for $\eta$ photoproduction. A stark
similarity, however, is that both formalism identify the polarization
observables as the prime candidates to investigate medium
modifications of the background processes such as the $D_{13}$
resonance. Meanwhile, as a consequence of the large dominance in the
differential cross section of the $S_{11}$ resonance, the quasifree
differential cross section provides an excellent tool to study medium
modifications to the $S_{11}$ resonance in order to distinquish
between various models that attempt to understand the $S_{11}$
resonance and its unique position as the lowest lying negative parity
state in the baryon spectrum. The basic message of this article is now
clear: to probe medium modifications to the $S_{11}$ resonance,
measure the differential cross section, to study the background
processes and their medium modifications, measure the polarization
observables.
 
\begin{acknowledgments} 
We are grateful to Professors M. Benmerrouche and A. J. Sarty for
providing us with their model for the elementary amplitude.
L. J. A. acknowledges the support of the Japan Society for the
Promotion of Science and the United States National Science Foundation
under award number 0002714.  G.C.H acknowledges financial support from
the Japanese Ministry of Education, Science and Technology for
research conducted at the Research Center for Nuclear Physics, Osaka,
Japan.  This material is based upon work supported by the National
Research Foundation under Grant numbers: GUN 2053786 (G.C.H), 2048567
(B.I.S.v.d.V).
\end{acknowledgments} 

\bibliography{eta} 

\begin{thebibliography}{55}
\expandafter\ifx\csname natexlab\endcsname\relax\def\natexlab#1{#1}\fi
\expandafter\ifx\csname bibnamefont\endcsname\relax
  \def\bibnamefont#1{#1}\fi
\expandafter\ifx\csname bibfnamefont\endcsname\relax
  \def\bibfnamefont#1{#1}\fi
\expandafter\ifx\csname citenamefont\endcsname\relax
  \def\citenamefont#1{#1}\fi
\expandafter\ifx\csname url\endcsname\relax
  \def\url#1{\texttt{#1}}\fi
\expandafter\ifx\csname urlprefix\endcsname\relax\def\urlprefix{URL }\fi
\providecommand{\bibinfo}[2]{#2}
\providecommand{\eprint}[2][]{\url{#2}}

\bibitem[{\citenamefont{Dover and Fishbane}(1990)}]{DF90}
\bibinfo{author}{\bibfnamefont{C.~B.} \bibnamefont{Dover}} \bibnamefont{and}
  \bibinfo{author}{\bibfnamefont{P.~M.} \bibnamefont{Fishbane}},
  \bibinfo{journal}{Phys. Rev. Lett.} \textbf{\bibinfo{volume}{64}},
  \bibinfo{pages}{3115} (\bibinfo{year}{1990}).

\bibitem[{\citenamefont{Krusche et~al.}(1995{\natexlab{a}})}]{krus95a}
\bibinfo{author}{\bibfnamefont{B.}~\bibnamefont{Krusche}} \bibnamefont{et~al.},
  \bibinfo{journal}{Phys. Rev. Lett.} \textbf{\bibinfo{volume}{74}},
  \bibinfo{pages}{3736} (\bibinfo{year}{1995}{\natexlab{a}}).

\bibitem[{\citenamefont{Krusche et~al.}(1995{\natexlab{b}})}]{krus95b}
\bibinfo{author}{\bibfnamefont{B.}~\bibnamefont{Krusche}} \bibnamefont{et~al.},
  \bibinfo{journal}{Phys. Lett.} \textbf{\bibinfo{volume}{B358}},
  \bibinfo{pages}{40} (\bibinfo{year}{1995}{\natexlab{b}}).

\bibitem[{\citenamefont{Benmerrouche and Mukhopadhyay}(1991)}]{BMN91}
\bibinfo{author}{\bibfnamefont{M.}~\bibnamefont{Benmerrouche}}
  \bibnamefont{and} \bibinfo{author}{\bibfnamefont{N.~C.}
  \bibnamefont{Mukhopadhyay}}, \bibinfo{journal}{Phys. Rev. Lett.}
  \textbf{\bibinfo{volume}{67}}, \bibinfo{pages}{1070} (\bibinfo{year}{1991}).

\bibitem[{\citenamefont{Li}(1995)}]{Li95}
\bibinfo{author}{\bibfnamefont{Z.-p.} \bibnamefont{Li}},
  \bibinfo{journal}{Phys. Rev.} \textbf{\bibinfo{volume}{D52}},
  \bibinfo{pages}{4961} (\bibinfo{year}{1995}),
  \eprint[http://arXiv.org/abs]{nucl-th/9506033}.

\bibitem[{\citenamefont{Benmerrouche et~al.}(1995)\citenamefont{Benmerrouche,
  Mukhopadhyay, and Zhang}}]{bmz95}
\bibinfo{author}{\bibfnamefont{M.}~\bibnamefont{Benmerrouche}},
  \bibinfo{author}{\bibfnamefont{N.~C.} \bibnamefont{Mukhopadhyay}},
  \bibnamefont{and} \bibinfo{author}{\bibfnamefont{J.~F.} \bibnamefont{Zhang}},
  \bibinfo{journal}{Phys. Rev.} \textbf{\bibinfo{volume}{D51}},
  \bibinfo{pages}{3237} (\bibinfo{year}{1995}),
  \eprint[http://arXiv.org/abs]{hep-ph/9412248}.

\bibitem[{\citenamefont{Kaiser et~al.}(1995)\citenamefont{Kaiser, Siegel, and
  Weise}}]{KNS95}
\bibinfo{author}{\bibfnamefont{N.}~\bibnamefont{Kaiser}},
  \bibinfo{author}{\bibfnamefont{P.~B.} \bibnamefont{Siegel}},
  \bibnamefont{and} \bibinfo{author}{\bibfnamefont{W.}~\bibnamefont{Weise}},
  \bibinfo{journal}{Phys. Lett.} \textbf{\bibinfo{volume}{B362}},
  \bibinfo{pages}{23} (\bibinfo{year}{1995}),
  \eprint[http://arXiv.org/abs]{nucl-th/9507036}.

\bibitem[{\citenamefont{Sauermann et~al.}(1995)\citenamefont{Sauermann, Friman,
  and Norenberg}}]{SFN95}
\bibinfo{author}{\bibfnamefont{C.}~\bibnamefont{Sauermann}},
  \bibinfo{author}{\bibfnamefont{B.~L.} \bibnamefont{Friman}},
  \bibnamefont{and}
  \bibinfo{author}{\bibfnamefont{W.}~\bibnamefont{Norenberg}},
  \bibinfo{journal}{Phys. Lett.} \textbf{\bibinfo{volume}{B341}},
  \bibinfo{pages}{261} (\bibinfo{year}{1995}),
  \eprint[http://arXiv.org/abs]{nucl-th/9408012}.

\bibitem[{\citenamefont{Jido et~al.}(1996)\citenamefont{Jido, Kodama, and
  Oka}}]{JKO96}
\bibinfo{author}{\bibfnamefont{D.}~\bibnamefont{Jido}},
  \bibinfo{author}{\bibfnamefont{N.}~\bibnamefont{Kodama}}, \bibnamefont{and}
  \bibinfo{author}{\bibfnamefont{M.}~\bibnamefont{Oka}},
  \bibinfo{journal}{Phys. Rev.} \textbf{\bibinfo{volume}{D54}},
  \bibinfo{pages}{4532} (\bibinfo{year}{1996}),
  \eprint[http://arXiv.org/abs]{hep-ph/9604280}.

\bibitem[{\citenamefont{Kim and Lee}(1997)}]{KL97}
\bibinfo{author}{\bibfnamefont{H.-c.} \bibnamefont{Kim}} \bibnamefont{and}
  \bibinfo{author}{\bibfnamefont{S.~H.} \bibnamefont{Lee}},
  \bibinfo{journal}{Phys. Rev.} \textbf{\bibinfo{volume}{D56}},
  \bibinfo{pages}{4278} (\bibinfo{year}{1997}),
  \eprint[http://arXiv.org/abs]{nucl-th/9704035}.

\bibitem[{\citenamefont{Lee and Kim}(1997)}]{LK97}
\bibinfo{author}{\bibfnamefont{S.~H.} \bibnamefont{Lee}} \bibnamefont{and}
  \bibinfo{author}{\bibfnamefont{H.-c.} \bibnamefont{Kim}},
  \bibinfo{journal}{Nucl. Phys.} \textbf{\bibinfo{volume}{A612}},
  \bibinfo{pages}{418} (\bibinfo{year}{1997}),
  \eprint[http://arXiv.org/abs]{nucl-th/9608038}.

\bibitem[{\citenamefont{Jido et~al.}(1998)\citenamefont{Jido, Oka, and
  Hosaka}}]{JOH98}
\bibinfo{author}{\bibfnamefont{D.}~\bibnamefont{Jido}},
  \bibinfo{author}{\bibfnamefont{M.}~\bibnamefont{Oka}}, \bibnamefont{and}
  \bibinfo{author}{\bibfnamefont{A.}~\bibnamefont{Hosaka}},
  \bibinfo{journal}{Phys. Rev. Lett.} \textbf{\bibinfo{volume}{80}},
  \bibinfo{pages}{448} (\bibinfo{year}{1998}),
  \eprint[http://arXiv.org/abs]{hep-ph/9707307}.

\bibitem[{\citenamefont{DeTar and Kunihiro}(1989)}]{DK89}
\bibinfo{author}{\bibfnamefont{C.}~\bibnamefont{DeTar}} \bibnamefont{and}
  \bibinfo{author}{\bibfnamefont{T.}~\bibnamefont{Kunihiro}},
  \bibinfo{journal}{Phys. Rev.} \textbf{\bibinfo{volume}{D39}},
  \bibinfo{pages}{2805} (\bibinfo{year}{1989}).

\bibitem[{\citenamefont{Jido et~al.}(2000)\citenamefont{Jido, Nemoto, Oka, and
  Hosaka}}]{JNOH00}
\bibinfo{author}{\bibfnamefont{D.}~\bibnamefont{Jido}},
  \bibinfo{author}{\bibfnamefont{Y.}~\bibnamefont{Nemoto}},
  \bibinfo{author}{\bibfnamefont{M.}~\bibnamefont{Oka}}, \bibnamefont{and}
  \bibinfo{author}{\bibfnamefont{A.}~\bibnamefont{Hosaka}},
  \bibinfo{journal}{Nucl. Phys.} \textbf{\bibinfo{volume}{A671}},
  \bibinfo{pages}{471} (\bibinfo{year}{2000}),
  \eprint[http://arXiv.org/abs]{hep-ph/9805306}.

\bibitem[{\citenamefont{Jido et~al.}(2001)\citenamefont{Jido, Oka, and
  Hosaka}}]{JOH01}
\bibinfo{author}{\bibfnamefont{D.}~\bibnamefont{Jido}},
  \bibinfo{author}{\bibfnamefont{M.}~\bibnamefont{Oka}}, \bibnamefont{and}
  \bibinfo{author}{\bibfnamefont{A.}~\bibnamefont{Hosaka}},
  \bibinfo{journal}{Prog. Theor. Phys.} \textbf{\bibinfo{volume}{106}},
  \bibinfo{pages}{873} (\bibinfo{year}{2001}),
  \eprint[http://arXiv.org/abs]{hep-ph/0110005}.

\bibitem[{\citenamefont{Kim et~al.}(1998)\citenamefont{Kim, Jido, and
  Oka}}]{KJO98}
\bibinfo{author}{\bibfnamefont{H.-c.} \bibnamefont{Kim}},
  \bibinfo{author}{\bibfnamefont{D.}~\bibnamefont{Jido}}, \bibnamefont{and}
  \bibinfo{author}{\bibfnamefont{M.}~\bibnamefont{Oka}},
  \bibinfo{journal}{Nucl. Phys.} \textbf{\bibinfo{volume}{A640}},
  \bibinfo{pages}{77} (\bibinfo{year}{1998}),
  \eprint[http://arXiv.org/abs]{hep-ph/9806275}.

\bibitem[{\citenamefont{Oka}()}]{oka}
\bibinfo{author}{\bibfnamefont{M.}~\bibnamefont{Oka}}, \bibinfo{note}{private
  communication}.

\bibitem[{\citenamefont{Piekarewicz et~al.}(1997)\citenamefont{Piekarewicz,
  Sarty, and Benmerrouche}}]{PSB97}
\bibinfo{author}{\bibfnamefont{J.}~\bibnamefont{Piekarewicz}},
  \bibinfo{author}{\bibfnamefont{A.~J.} \bibnamefont{Sarty}}, \bibnamefont{and}
  \bibinfo{author}{\bibfnamefont{M.}~\bibnamefont{Benmerrouche}},
  \bibinfo{journal}{Phys. Rev.} \textbf{\bibinfo{volume}{C55}},
  \bibinfo{pages}{2571} (\bibinfo{year}{1997}),
  \eprint[http://arXiv.org/abs]{nucl-th/9701019}.

\bibitem[{\citenamefont{Peters et~al.}(1998)\citenamefont{Peters, Lenske, and
  Mosel}}]{PLM98}
\bibinfo{author}{\bibfnamefont{W.}~\bibnamefont{Peters}},
  \bibinfo{author}{\bibfnamefont{H.}~\bibnamefont{Lenske}}, \bibnamefont{and}
  \bibinfo{author}{\bibfnamefont{U.}~\bibnamefont{Mosel}},
  \bibinfo{journal}{Nucl. Phys.} \textbf{\bibinfo{volume}{A642}},
  \bibinfo{pages}{506} (\bibinfo{year}{1998}),
  \eprint[http://arXiv.org/abs]{nucl-th/9807002}.

\bibitem[{\citenamefont{Abu-Raddad et~al.}(1999)\citenamefont{Abu-Raddad,
  Piekarewicz, Sarty, and Rego}}]{APSR99}
\bibinfo{author}{\bibfnamefont{L.~J.} \bibnamefont{Abu-Raddad}},
  \bibinfo{author}{\bibfnamefont{J.}~\bibnamefont{Piekarewicz}},
  \bibinfo{author}{\bibfnamefont{A.~J.} \bibnamefont{Sarty}}, \bibnamefont{and}
  \bibinfo{author}{\bibfnamefont{R.~A.} \bibnamefont{Rego}},
  \bibinfo{journal}{Phys. Rev.} \textbf{\bibinfo{volume}{C60}},
  \bibinfo{pages}{054606} (\bibinfo{year}{1999}),
  \eprint[http://arXiv.org/abs]{nucl-th/9812061}.

\bibitem[{\citenamefont{Bennhold and Tanabe}(1990)}]{bt90}
\bibinfo{author}{\bibfnamefont{C.}~\bibnamefont{Bennhold}} \bibnamefont{and}
  \bibinfo{author}{\bibfnamefont{H.}~\bibnamefont{Tanabe}},
  \bibinfo{journal}{Phys. Lett.} \textbf{\bibinfo{volume}{B243}},
  \bibinfo{pages}{13} (\bibinfo{year}{1990}).

\bibitem[{\citenamefont{Halderson and Rosenthal}(1990)}]{HR90}
\bibinfo{author}{\bibfnamefont{D.}~\bibnamefont{Halderson}} \bibnamefont{and}
  \bibinfo{author}{\bibfnamefont{A.~S.} \bibnamefont{Rosenthal}},
  \bibinfo{journal}{Phys. Rev.} \textbf{\bibinfo{volume}{C42}},
  \bibinfo{pages}{2584} (\bibinfo{year}{1990}).

\bibitem[{\citenamefont{Carrasco}(1993)}]{Carrasco93}
\bibinfo{author}{\bibfnamefont{R.~C.} \bibnamefont{Carrasco}},
  \bibinfo{journal}{Phys. Rev.} \textbf{\bibinfo{volume}{C48}},
  \bibinfo{pages}{2333} (\bibinfo{year}{1993}).

\bibitem[{\citenamefont{Chen and Chiang}(1994)}]{CC94}
\bibinfo{author}{\bibfnamefont{L.}~\bibnamefont{Chen}} \bibnamefont{and}
  \bibinfo{author}{\bibfnamefont{H.-C.} \bibnamefont{Chiang}},
  \bibinfo{journal}{Phys. Lett.} \textbf{\bibinfo{volume}{B329}},
  \bibinfo{pages}{424} (\bibinfo{year}{1994}).

\bibitem[{\citenamefont{Tiator et~al.}(1994)\citenamefont{Tiator, Bennhold, and
  Kamalov}}]{TBK94}
\bibinfo{author}{\bibfnamefont{L.}~\bibnamefont{Tiator}},
  \bibinfo{author}{\bibfnamefont{C.}~\bibnamefont{Bennhold}}, \bibnamefont{and}
  \bibinfo{author}{\bibfnamefont{S.~S.} \bibnamefont{Kamalov}},
  \bibinfo{journal}{Nucl. Phys.} \textbf{\bibinfo{volume}{A580}},
  \bibinfo{pages}{455} (\bibinfo{year}{1994}), \eprint{nucl-th/9404013}.

\bibitem[{\citenamefont{Barshay and Bramon}(1994)}]{BB94}
\bibinfo{author}{\bibfnamefont{S.}~\bibnamefont{Barshay}} \bibnamefont{and}
  \bibinfo{author}{\bibfnamefont{A.}~\bibnamefont{Bramon}},
  \bibinfo{journal}{Mod. Phys. Lett.} \textbf{\bibinfo{volume}{A9}},
  \bibinfo{pages}{1727} (\bibinfo{year}{1994}).

\bibitem[{\citenamefont{Hombach et~al.}(1995)\citenamefont{Hombach, Engel,
  Teis, and Mosel}}]{HETM95}
\bibinfo{author}{\bibfnamefont{A.}~\bibnamefont{Hombach}},
  \bibinfo{author}{\bibfnamefont{A.}~\bibnamefont{Engel}},
  \bibinfo{author}{\bibfnamefont{S.}~\bibnamefont{Teis}}, \bibnamefont{and}
  \bibinfo{author}{\bibfnamefont{U.}~\bibnamefont{Mosel}}, \bibinfo{journal}{Z.
  Phys.} \textbf{\bibinfo{volume}{A352}}, \bibinfo{pages}{223}
  (\bibinfo{year}{1995}), \eprint[http://arXiv.org/abs]{nucl-th/9411025}.

\bibitem[{\citenamefont{Lee et~al.}(1996)\citenamefont{Lee, Wright, Bennhold,
  and Tiator}}]{lwbt96}
\bibinfo{author}{\bibfnamefont{F.~X.} \bibnamefont{Lee}},
  \bibinfo{author}{\bibfnamefont{L.~E.} \bibnamefont{Wright}},
  \bibinfo{author}{\bibfnamefont{C.}~\bibnamefont{Bennhold}}, \bibnamefont{and}
  \bibinfo{author}{\bibfnamefont{L.}~\bibnamefont{Tiator}},
  \bibinfo{journal}{Nucl. Phys.} \textbf{\bibinfo{volume}{A603}},
  \bibinfo{pages}{345} (\bibinfo{year}{1996}),
  \eprint[http://arXiv.org/abs]{nucl-th/9601001}.

\bibitem[{\citenamefont{Benmerrouche}(1992)}]{benm92}
\bibinfo{author}{\bibfnamefont{M.}~\bibnamefont{Benmerrouche}}, Ph.D. thesis,
  \bibinfo{school}{Rensselaer Polytechnic Institute} (\bibinfo{year}{1992}).

\bibitem[{\citenamefont{Mukhopadhyay et~al.}(1995)\citenamefont{Mukhopadhyay,
  Zhang, and Benmerrouche}}]{mzb95}
\bibinfo{author}{\bibfnamefont{N.~C.} \bibnamefont{Mukhopadhyay}},
  \bibinfo{author}{\bibfnamefont{J.~F.} \bibnamefont{Zhang}}, \bibnamefont{and}
  \bibinfo{author}{\bibfnamefont{M.}~\bibnamefont{Benmerrouche}},
  \bibinfo{journal}{Phys. Lett.} \textbf{\bibinfo{volume}{B364}},
  \bibinfo{pages}{1} (\bibinfo{year}{1995}),
  \eprint[http://arXiv.org/abs]{hep-ph/9510307}.

\bibitem[{\citenamefont{Abu-Raddad}(2000)}]{thesis}
\bibinfo{author}{\bibfnamefont{L.~J.} \bibnamefont{Abu-Raddad}}, Ph.D. thesis,
  \bibinfo{school}{Florida State University} (\bibinfo{year}{2000}),
  \eprint[http://arXiv.org/abs]{nucl-th/0005068}.

\bibitem[{\citenamefont{Abu-Raddad and Piekarewicz}(2000)}]{AP00}
\bibinfo{author}{\bibfnamefont{L.~J.} \bibnamefont{Abu-Raddad}}
  \bibnamefont{and}
  \bibinfo{author}{\bibfnamefont{J.}~\bibnamefont{Piekarewicz}},
  \bibinfo{journal}{Phys. Rev.} \textbf{\bibinfo{volume}{C61}},
  \bibinfo{pages}{014604} (\bibinfo{year}{2000}),
  \eprint[http://arXiv.org/abs]{nucl-th/9906066}.

\bibitem[{\citenamefont{Abu-Raddad and Piekarewicz}(2001)}]{AP01}
\bibinfo{author}{\bibfnamefont{L.~J.} \bibnamefont{Abu-Raddad}}
  \bibnamefont{and}
  \bibinfo{author}{\bibfnamefont{J.}~\bibnamefont{Piekarewicz}},
  \bibinfo{journal}{Phys. Rev.} \textbf{\bibinfo{volume}{C64}},
  \bibinfo{pages}{064616} (\bibinfo{year}{2001}),
  \eprint[http://arXiv.org/abs]{nucl-th/0106015}.

\bibitem[{\citenamefont{Li et~al.}(1993)\citenamefont{Li, Wright, and
  Bennhold}}]{lwb93}
\bibinfo{author}{\bibfnamefont{X.}~\bibnamefont{Li}},
  \bibinfo{author}{\bibfnamefont{L.~E.} \bibnamefont{Wright}},
  \bibnamefont{and} \bibinfo{author}{\bibfnamefont{C.}~\bibnamefont{Bennhold}},
  \bibinfo{journal}{Phys. Rev.} \textbf{\bibinfo{volume}{C48}},
  \bibinfo{pages}{816} (\bibinfo{year}{1993}).

\bibitem[{\citenamefont{Bennhold et~al.}(1998)\citenamefont{Bennhold, Lee,
  Mart, and Wright}}]{blmw98}
\bibinfo{author}{\bibfnamefont{C.}~\bibnamefont{Bennhold}},
  \bibinfo{author}{\bibfnamefont{F.~X.} \bibnamefont{Lee}},
  \bibinfo{author}{\bibfnamefont{T.}~\bibnamefont{Mart}}, \bibnamefont{and}
  \bibinfo{author}{\bibfnamefont{L.~E.} \bibnamefont{Wright}},
  \bibinfo{journal}{Nucl. Phys.} \textbf{\bibinfo{volume}{A639}},
  \bibinfo{pages}{227c} (\bibinfo{year}{1998}),
  \eprint[http://arXiv.org/abs]{nucl-th/9712075}.

\bibitem[{\citenamefont{Gardner and Piekarewicz}(1994)}]{gp94}
\bibinfo{author}{\bibfnamefont{S.}~\bibnamefont{Gardner}} \bibnamefont{and}
  \bibinfo{author}{\bibfnamefont{J.}~\bibnamefont{Piekarewicz}},
  \bibinfo{journal}{Phys. Rev.} \textbf{\bibinfo{volume}{C50}},
  \bibinfo{pages}{2822} (\bibinfo{year}{1994}),
  \eprint[http://arXiv.org/abs]{nucl-th/9401001}.

\bibitem[{\citenamefont{Caballero et~al.}(1998)\citenamefont{Caballero,
  Donnelly, Moya~de Guerra, and Udias}}]{cdmu98}
\bibinfo{author}{\bibfnamefont{J.~A.} \bibnamefont{Caballero}},
  \bibinfo{author}{\bibfnamefont{T.~W.} \bibnamefont{Donnelly}},
  \bibinfo{author}{\bibfnamefont{E.}~\bibnamefont{Moya~de Guerra}},
  \bibnamefont{and} \bibinfo{author}{\bibfnamefont{J.~M.} \bibnamefont{Udias}},
  \bibinfo{journal}{Nucl. Phys.} \textbf{\bibinfo{volume}{A632}},
  \bibinfo{pages}{323} (\bibinfo{year}{1998}),
  \eprint[http://arXiv.org/abs]{nucl-th/9710038}.

\bibitem[{\citenamefont{Hejny et~al.}(1999)}]{Hejny99}
\bibinfo{author}{\bibfnamefont{V.}~\bibnamefont{Hejny}} \bibnamefont{et~al.},
  \bibinfo{journal}{Eur. Phys. J.} \textbf{\bibinfo{volume}{A6}},
  \bibinfo{pages}{83} (\bibinfo{year}{1999}).

\bibitem[{\citenamefont{Hejny et~al.}(2001)}]{Weiss_EJoPA13_2001}
\bibinfo{author}{\bibfnamefont{V.}~\bibnamefont{Hejny}} \bibnamefont{et~al.},
  \bibinfo{journal}{Eur. Phys. J.} \textbf{\bibinfo{volume}{A13}},
  \bibinfo{pages}{493} (\bibinfo{year}{2001}).

\bibitem[{\citenamefont{Robig-Landau et~al.}(1996)}]{RobigLandau_PLB373_1996}
\bibinfo{author}{\bibfnamefont{M.}~\bibnamefont{Robig-Landau}}
  \bibnamefont{et~al.}, \bibinfo{journal}{Phys. Lett.}
  \textbf{\bibinfo{volume}{B373}}, \bibinfo{pages}{45} (\bibinfo{year}{1996}).

\bibitem[{\citenamefont{Mandl and Shaw}(1984)}]{MS84}
\bibinfo{author}{\bibfnamefont{F.}~\bibnamefont{Mandl}} \bibnamefont{and}
  \bibinfo{author}{\bibfnamefont{G.}~\bibnamefont{Shaw}},
  \emph{\bibinfo{title}{Quantum Field Theory}}, Wiley-Interscience Publication
  (\bibinfo{publisher}{Wiley}, \bibinfo{address}{Chichester, UK},
  \bibinfo{year}{1984}), \bibinfo{note}{p 138}.

\bibitem[{\citenamefont{Williams et~al.}(1990)\citenamefont{Williams, Ji, and
  Cotanch}}]{wcc90}
\bibinfo{author}{\bibfnamefont{R.}~\bibnamefont{Williams}},
  \bibinfo{author}{\bibfnamefont{C.~R.} \bibnamefont{Ji}}, \bibnamefont{and}
  \bibinfo{author}{\bibfnamefont{S.~R.} \bibnamefont{Cotanch}},
  \bibinfo{journal}{Phys. Rev.} \textbf{\bibinfo{volume}{D41}},
  \bibinfo{pages}{1449} (\bibinfo{year}{1990}).

\bibitem[{\citenamefont{Nagl et~al.}(1991)\citenamefont{Nagl, Devanathan, and
  {\"U}berall}}]{ndu91}
\bibinfo{author}{\bibfnamefont{A.}~\bibnamefont{Nagl}},
  \bibinfo{author}{\bibfnamefont{V.}~\bibnamefont{Devanathan}},
  \bibnamefont{and}
  \bibinfo{author}{\bibfnamefont{H.}~\bibnamefont{{\"U}berall}},
  \emph{\bibinfo{title}{Nuclear Pion Photoproduction}}, vol.
  \bibinfo{volume}{120} of \emph{\bibinfo{series}{Springer Tracts in Modern
  Physics}} (\bibinfo{publisher}{Springer-Verlag}, \bibinfo{address}{Berlin
  Heidelberg, Germany}, \bibinfo{year}{1991}).

\bibitem[{\citenamefont{Chew et~al.}(1957)\citenamefont{Chew, Goldberger, Low,
  and Nambu}}]{cgln57}
\bibinfo{author}{\bibfnamefont{G.~F.} \bibnamefont{Chew}},
  \bibinfo{author}{\bibfnamefont{M.~L.} \bibnamefont{Goldberger}},
  \bibinfo{author}{\bibfnamefont{F.~E.} \bibnamefont{Low}}, \bibnamefont{and}
  \bibinfo{author}{\bibfnamefont{Y.}~\bibnamefont{Nambu}},
  \bibinfo{journal}{Phys. Rev.} \textbf{\bibinfo{volume}{106}},
  \bibinfo{pages}{1345} (\bibinfo{year}{1957}).

\bibitem[{\citenamefont{Bennhold and Tanabe}(1991)}]{bt91}
\bibinfo{author}{\bibfnamefont{C.}~\bibnamefont{Bennhold}} \bibnamefont{and}
  \bibinfo{author}{\bibfnamefont{H.}~\bibnamefont{Tanabe}},
  \bibinfo{journal}{Nucl. Phys.} \textbf{\bibinfo{volume}{A530}},
  \bibinfo{pages}{625} (\bibinfo{year}{1991}).

\bibitem[{\citenamefont{Homma et~al.}(1988)\citenamefont{Homma, Kanazawa,
  Maruyama, Murata, Okuno, Sasaki, and Taniguchi}}]{Homma88}
\bibinfo{author}{\bibfnamefont{S.}~\bibnamefont{Homma}},
  \bibinfo{author}{\bibfnamefont{M.}~\bibnamefont{Kanazawa}},
  \bibinfo{author}{\bibfnamefont{K.}~\bibnamefont{Maruyama}},
  \bibinfo{author}{\bibfnamefont{Y.}~\bibnamefont{Murata}},
  \bibinfo{author}{\bibfnamefont{H.}~\bibnamefont{Okuno}},
  \bibinfo{author}{\bibfnamefont{H.}~\bibnamefont{Sasaki}}, \bibnamefont{and}
  \bibinfo{author}{\bibfnamefont{T.}~\bibnamefont{Taniguchi}},
  \bibinfo{journal}{J. Phys. Soc. Jap.} \textbf{\bibinfo{volume}{57}},
  \bibinfo{pages}{828} (\bibinfo{year}{1988}).

\bibitem[{\citenamefont{Hicks et~al.}(1973)\citenamefont{Hicks, Deans, Jacobs,
  Lyons, and Montgomery}}]{Hicks73}
\bibinfo{author}{\bibfnamefont{H.~R.} \bibnamefont{Hicks}},
  \bibinfo{author}{\bibfnamefont{S.~R.} \bibnamefont{Deans}},
  \bibinfo{author}{\bibfnamefont{D.~T.} \bibnamefont{Jacobs}},
  \bibinfo{author}{\bibfnamefont{P.~W.} \bibnamefont{Lyons}}, \bibnamefont{and}
  \bibinfo{author}{\bibfnamefont{D.~L.} \bibnamefont{Montgomery}},
  \bibinfo{journal}{Phys. Rev.} \textbf{\bibinfo{volume}{D7}},
  \bibinfo{pages}{2614} (\bibinfo{year}{1973}), \bibinfo{note}{and references
  therein}.

\bibitem[{\citenamefont{Serot and Walecka}(1986)}]{SW86}
\bibinfo{author}{\bibfnamefont{B.~D.} \bibnamefont{Serot}} \bibnamefont{and}
  \bibinfo{author}{\bibfnamefont{J.~D.} \bibnamefont{Walecka}},
  \bibinfo{journal}{Adv. Nucl. Phys.} \textbf{\bibinfo{volume}{16}},
  \bibinfo{pages}{1} (\bibinfo{year}{1986}).

\bibitem[{\citenamefont{Sakurai}(1973)}]{sak73}
\bibinfo{author}{\bibfnamefont{J.}~\bibnamefont{Sakurai}},
  \emph{\bibinfo{title}{Advanced Quantum Mechanics}}
  (\bibinfo{publisher}{Addison-Wesley}, \bibinfo{year}{1973}).

\bibitem[{\citenamefont{Taylor}(1972)}]{Taylor72}
\bibinfo{author}{\bibfnamefont{J.~R.} \bibnamefont{Taylor}},
  \emph{\bibinfo{title}{Scattering Theory: The Quantum Theory of
  Nonrelativistic Collisions}} (\bibinfo{publisher}{John Wiley \& Sons},
  \bibinfo{year}{1972}).

\bibitem[{\citenamefont{Mertig et~al.}(1991)\citenamefont{Mertig, Bohm, and
  Denner}}]{MBD91}
\bibinfo{author}{\bibfnamefont{R.}~\bibnamefont{Mertig}},
  \bibinfo{author}{\bibfnamefont{M.}~\bibnamefont{Bohm}}, \bibnamefont{and}
  \bibinfo{author}{\bibfnamefont{A.}~\bibnamefont{Denner}},
  \bibinfo{journal}{Comput. Phys. Commun.} \textbf{\bibinfo{volume}{64}},
  \bibinfo{pages}{345} (\bibinfo{year}{1991}),
  \bibinfo{note}{http://www.feyncalc.org/}.

\bibitem[{\citenamefont{De~Baenst}(1970)}]{DeBaenst70}
\bibinfo{author}{\bibfnamefont{P.}~\bibnamefont{De~Baenst}},
  \bibinfo{journal}{Nucl. Phys.} \textbf{\bibinfo{volume}{B24}},
  \bibinfo{pages}{633} (\bibinfo{year}{1970}).

\bibitem[{\citenamefont{Bernard et~al.}(1992)\citenamefont{Bernard, Kaiser, and
  Meissner}}]{BKM92}
\bibinfo{author}{\bibfnamefont{V.}~\bibnamefont{Bernard}},
  \bibinfo{author}{\bibfnamefont{N.}~\bibnamefont{Kaiser}}, \bibnamefont{and}
  \bibinfo{author}{\bibfnamefont{U.~G.} \bibnamefont{Meissner}},
  \bibinfo{journal}{Nucl. Phys.} \textbf{\bibinfo{volume}{B383}},
  \bibinfo{pages}{442} (\bibinfo{year}{1992}).

\bibitem[{\citenamefont{Gell-Mann et~al.}(1968)\citenamefont{Gell-Mann, Oakes,
  and Renner}}]{GMOR68}
\bibinfo{author}{\bibfnamefont{M.}~\bibnamefont{Gell-Mann}},
  \bibinfo{author}{\bibfnamefont{R.~J.} \bibnamefont{Oakes}}, \bibnamefont{and}
  \bibinfo{author}{\bibfnamefont{B.}~\bibnamefont{Renner}},
  \bibinfo{journal}{Phys. Rev.} \textbf{\bibinfo{volume}{175}},
  \bibinfo{pages}{2195} (\bibinfo{year}{1968}).

\bibitem[{\citenamefont{Manohar and Georgi}(1984)}]{MG84}
\bibinfo{author}{\bibfnamefont{A.}~\bibnamefont{Manohar}} \bibnamefont{and}
  \bibinfo{author}{\bibfnamefont{H.}~\bibnamefont{Georgi}},
  \bibinfo{journal}{Nucl. Phys.} \textbf{\bibinfo{volume}{B234}},
  \bibinfo{pages}{189} (\bibinfo{year}{1984}).

\end{thebibliography}

\end{document}